\pdfoutput=1
\documentclass[11pt]{article}

\usepackage{color}
\usepackage{todonotes}
\usepackage{verbatim}
\usepackage{booktabs}
\usepackage{graphicx}
\usepackage[colorlinks=true,linkcolor=black,citecolor=black]{hyperref}
\usepackage{epstopdf}
\usepackage[toc, page]{appendix}
\usepackage{subfigure}
\usepackage{chngcntr}
\usepackage[margin=2cm]{geometry}

\usepackage{palatino}

\def\affilnum#1{${}^{#1}$}
\def\affil#1{${}^{#1}$}

\def\corraddr#1{\footnotetext[1]{Correspondence to: #1\stepcounter{footnote}}}
\def\keywords#1{\small{KEY WORDS:}{#1}}
\def\corrauth{\footnotemark[1]}
\RequirePackage{graphicx}
\RequirePackage{pifont,latexsym,ifthen,rotating,calc,textcase,booktabs,color}
\RequirePackage{amsfonts,amssymb,amsbsy,amsmath,amsthm}
\RequirePackage[errorshow]{tracefnt}

\title{\textbf{Methods and Metrics for Fair Server Assessment \\ under Real-Time Financial Workloads}}

\begin{document}
\typeout{  >>>> Start processing document }
\date{}

\author{Giorgis Georgakoudis\affil{1}\corrauth,\ 
        Charles J. Gillan\affil{2},
        Ahmed Sayed\affil{1}, \\
        Ivor Spence\affil{1},
        Richard Faloon\affil{3},
        and Dimitrios S. Nikolopoulos\affil{1}}

\maketitle

\begin{center}
\affilnum{1}The School of Electronics, Electrical Engineering and Computer Science,
                  Queen's University Belfast, Northern Ireland BT7 1NN, United Kingdom \break
\affilnum{2}The Institute for Electronics Communications and Information Technology, 
                 Queen's University Belfast, The Northern Ireland Science Park
                  Queen's Road, Belfast, Northern Ireland BT3 9DT, United Kingdom  \break
\affilnum{3}Neueda Consulting Limited, Glenwood Business Centre, Springbank Industrial Estate
               Belfast, Northern Ireland BT17 0QL, United Kingdom
\end{center}

\corraddr{The School of Electronics, Electrical Engineering and Computer Science,
          Queen's University Belfast, Northern Ireland BT7 1NN, United Kingdom}


%
%
   %
\begin{abstract}
Energy efficiency has been a daunting challenge for datacenters.
The financial industry operates some of the largest datacenters in the world.
With increasing energy costs and the financial services sector growth, 
emerging financial analytics workloads may incur 
extremely high operational costs, to meet 
their latency targets.
Microservers have recently emerged as an alternative 
to high-end servers, promising scalable performance and low energy consumption in datacenters via scale-out. 
Unfortunately, stark differences in architectural features, form factor and design considerations
make a fair comparison between servers and microservers exceptionally challenging.
\par
In this paper we present a rigorous methodology and new metrics for fair comparison of server and 
microserver platforms.  We deploy our methodology and metrics to compare a microserver with ARM
cores against two servers with x86 cores, running the same real-time financial analytics workload. 
We define workload-specific but platform-independent performance metrics for platform 
comparison, targeting both datacenter operators and end users.
\par
Our methodology establishes that a server based the Xeon Phi processor delivers the highest 
performance and energy-efficiency. However, by scaling out energy-efficient microservers, we 
achieve competitive or better energy-efficiency than a power-equivalent server with two Sandy Bridge sockets, 
despite the microserver's slower cores.  Using a new iso-QoS (iso-Quality of Service) metric,
we find that the ARM microserver scales enough to meet 
market throughput demand, i.e.\ a 100\% QoS in terms of timely option pricing,
with as little as 55\% of the energy consumed by the Sandy Bridge server. 
\par
   %
\end{abstract}
   %
%
%

\section{Introduction}

Financial analytics are one of the most important services in modern capital markets. Real-time financial
analytics, such as option pricing and trading risk assessment incur a high premium on top of basic
service costs, due to their latency criticality. 

Financial service providers operate datacenters which are co-located with market data feeds, to meet
the low latency requirements of real-time analytics. These datacenters incur extraordinarily high cost of ownership
due to hardware overprovisioning, their scale, and their carbon footprint. The choice of compute server architecture for
these datacenters fundamentally dictates their cost of ownership and sustainability. 
Unfortunately, the financial industry lacks concrete methodologies and metrics for fair assessment of
server architectures as computational workhorses for emerging analytics workloads.

In this paper we present a rigorous methodology and a set of metrics for evaluation of
server performance and energy efficiency under real-time financial analytics workloads.
We base our work on an earlier contribution which explored the viability of an ARM microserver
in comparison to Intel Sandy Bridge servers for real-time option pricing 
workloads~\cite{Gillan:2014:VMF:2688424.2688429}.
We introduce a platform-independent workload setup and optimization methodology, in conjunction with
platform-independent metrics of performance, energy-efficiency and quality of service.
We leverage these tools to fairly and thoroughly compare three server architectures that represent vastly 
different price, cost, performance and power trade-offs: a Calxeda  ECX-1000 microserver~\cite{Novakovic:2014:SN:2541940.2541965} based on ARM Cortex A9
cores;  a  Dell server based on Intel Sandy Bridge cores; and a Dell server based on an Intel Xeon Phi 
co-processor. 

Our methodology yields important and often counter-intuitive findings. We establish that a server based the 
Xeon Phi processor delivers the highest performance and energy-efficiency. However, by scaling out 
energy-efficient ARM microservers within a 2U rack-mounted unit, we achieve competitive or better 
energy-efficiency than a power-equivalent server with two Intel Sandy Bridge sockets, 
despite the microserver's slower cores.  Using a new iso-QoS (iso-Quality of Service) metric,
we find that the ARM microserver scales enough to meet 
market throughput demand, i.e.\ a 100\% QoS in terms of timely option pricing,
with as little as 55\% of the energy consumed by the Sandy Bridge server. 

Using the same C and SIMD code basis, we uncover several more findings of interest. The servers
exhibit vast differences in energy-efficiency due to differences in the implementation of transcendental 
functions in hardware, as well as the implementation of vector units and ISAs. We find that power saving 
modes employing DVFS are invariably less energy-efficient than performance boosting modes for financial
analytics workloads.  We also find that while servers exhibit good
performance scaling with more cores, they also exhibit non-ideal energy scaling, thus providing
an opportunity for energy conservation via throttling concurrency.

The paper begins by providing background on microservers and server comparison methodologies 
(Section~\ref{sec:background}). We proceed by briefly defining financial option contracts and explaining the 
computational methods that we have used to obtain contract prices from spot prices in our
real-time workloads (Section~\ref{sec:option_pricing}).
We discuss common code optimization and vectorization methodology 
(Section~\ref{sec:Optimization Methodology}).
We move on to details of the platforms which we used for our
study and present our experimental and measurement methodology and metrics 
(Section~\ref{sec:Experimental Setup and Measurement Methodology}).
We proceed by presenting results from standalone kernel experiments using 
an actual stock market feed (Section~\ref{sec:experimentsandresults}).
We delve into a discussion of our study findings.  We then present and analyze
results from production-strength experiments (Section~\ref{sec:experimentsandresults2}). 
We discuss related work (Section~\ref{sec:related}) 
and finish the paper with summarizing our conclusions (Section~\ref{sec:conclusions}).

\section{Background}
\label{sec:background}

Microservers have the potential to reduce the energy consumption of datacenters by 
using low-power processors designed for the embedded systems market~\cite{Rajovic:2013:SCC:2503210.2503281}. 
Back-of-the-envelope estimates suggest that microserver processors such as the Samsung Exynos 
and Calxeda Highbank ECX-1000~\cite{Novakovic:2014:SN:2541940.2541965}
consume 5 to 30 times less power than high-end server-class counterparts, such as Intel's Sandy Bridge 
and are 5 to 15 times more energy-efficient in integer performance per Watt~\cite{Tudor:2013:UEC:2465529.2465553,Ou:2012:ECA:2310096.2310142}.  However, microserver processors are limited in floating point performance and SIMD/SIMT acceleration 
capabilities~\cite{Rajovic:2013:SCC:2503210.2503281}. 

In comparing servers to microservers, one must find a fair and unbiased methodology. A challenge lies in that
a standalone microserver represents a vastly different price, performance and operating power point than
a standalone server. 

\subsection{Metric considerations}

We stipulate that a comparison methodology should be platform neutral and ascertain that 
the server architectures under consideration meet the same, tangible target, be it performance, Quality of Service (QoS),  power consumption, or energy consumption. We thus allow scale-out or scale-up of each server
to meet the set target, while comparing other metrics of interest to rank the servers. 
Furthermore, the metrics used should reflect workload fluctuation, but not be unduly influenced by non-deterministic
environmental, hardware or system artifacts. Finally, we stipulate that any comparison should be based on the
same code base, developed with similar coding effort on each server. We elaborate on this issue in the following section. 

\subsection{Workload coding considerations}

We use real-time option pricing workloads, which represent one of the most important and performance-critical
domains of financial analytics. 
Option pricing is an inherently parallel problem, as each option contract is defined
over a single stock, independently. A portable implementation with similar effort across servers
described in our earlier work~\cite{Gillan:2014:VMF:2688424.2688429},
exploits parallelism in the workload by using UDP multicast channels to distribute prices and POSIX 
threads to distribute option pricing instances between cores and their hardware contexts, if present. 

In addition we use the same code base to exploit the vector processing units available on each server and
further accelerate the numerical calculations in our workloads. Our code base uses manual loop unrolling and vectorization using pragmas. 
Our implementation is expressed in the C language, which is the norm in computational finance. The C language standard presents a number of challenges for implicit 
vectorization by a compiler~\cite{Allen:1988:CCV:960116.54014}. Principal among these challenges are the 
facts that C imposes no constraints on subroutine argument aliasing, the ``for'' statement permits loop 
bounds and increment to change within the loop and loop operations are often implemented with pointers, 
which could be aliased or point to overlapping areas of memory. We demonstrate significant differences
in auto-vectorization capabilities between compilers. These differences render autovectorization inappropriate
for fair server comparison, even though on occasion (e.g.\ ICC on the Xeon Phi), autovectorization produces excellent results.

On the other hand, C compilers provide implicit vectorization with associated pragmas and command line 
options to control the vectorization process. These features help the authoring of portable SIMD code and
we leverage them in this work.  Still, there is no common standard for compilers to abide by to provide the 
same quality of vectorization nor is there a method to enforce different compilers to produce comparable
vectorization output. We offer insights on how to address this problem in our methodology. Manual 
vectorization has been extensively used in production environments~\cite{larson:Exploiting,shin:superword,shin:2007,mullan:1998,gillan:high} but remains heavily dependent on programmer skill.

\section{Computing Option Prices}
\label{sec:option_pricing}

In finance, the term Option means a derivative product which is a contract giving the holder the right 
to either buy (Call option) or to sell (Put option) one or more underlying assets, such as a fixed number 
of shares in a company, for a defined price and either on or before a contract end date. An Options
contract, unlike a Futures contract, does not impose an obligation on the holder to exercise 
their right. There are several types of Options distinguished by the terms in their contract. 
We construct real-time analytics workloads that continuously execute Monte Carlo or Binomial Tree option pricing models. These can price both the so-called exotic 
American or Bermudan options as well as European options. We focus 
on European options for which the Black-Scholes equation provides a 
closed form solution, against which our results can be compared for 
accuracy.

Black and Scholes~\cite{blackscholes:1972,blackscholes:1973} proposed a second-order partial differential 
equation which models the variation of the price of a European vanilla option, contractual strike price $P$, 
over time in years $T$, assuming that:

\begin{itemize}
 \item the underlying asset price (spot price), $S$ follows a log normal distribution,
 \item the volatility $\sigma$ of $S$ is constant and
 \item the risk free interest rate, or rate of return, $r$, expressed with continuous compounding is constant.
\end{itemize}

Under these conditions an analytic solution is:
\begin{equation}
   {\rm Price} = (-1)^{p} \left  ( 
                                  S N ((-1)^{p} d_1) - P {\rm e}^{-rT} N ((-1)^{p} d_2) 
                          \right )
   \label{eqn:blackscholes}
\end{equation}   

In this equation $p=1$ for a Call option and $p=0$ for a Put option and $d_1$ and $d_2$ are defined by:

\begin{equation}
   d_1 = \frac{1}{\sigma \sqrt{T}} 
                      \left  (
                              \log \left ( \frac{S}{N} \right  ) 
                               +  \frac{r + \sigma^2}{2} \; T
                      \right )
   \;\;\;\; {\rm and} \;\;\;\;
   d_2 = d_1 - \sigma\sqrt{T} 
   \label{eqn:blackscholesd}   
\end{equation}      

where $N(x)$ is the cumulative normal distribution function (CDF). 

Moving beyond the limitations of the Black-Scholes model, the price of an option on any given date can 
be modeled using stochastic calculus, essentially simulating the path of the underlying variables 
over a set of paths within a time window. Analytical solutions for the stochastic equations are not 
generally possible so that a variety of  computational numerical solution methods have been developed. 
European vanilla options can also be priced using these numerical methods. We apply two models, 
with distinct computational characteristics, as described next.

\subsection{Monte Carlo (MC)}

At the center of a Monte Carlo simulation of option pricing is a rate-limiting for-loop, an aspect that it shares in 
common with HPC applications in many fields. For a Put option the formula is~\cite{boyle:montecarlo}:

\begin{equation}
   {\rm Price} = 
   \frac{ {\rm e}^{-rT} }{N} \sum_{i=1}^{N} \max \left  (
                                                       0, S - P {\rm e}^{ 
                                                                         ( r - \frac{\sigma^2}{2} ) T + \sigma \sqrt{T} x_i 
                                                                        }
                                               \right ) 
   \label{eqn:montecarlo}                                            
\end{equation}

In equation~\ref{eqn:montecarlo}, ${x_i\ (i=1\ldots N)}$ are a set of random numbers drawn from the 
standard normal distribution. The formula for a Call option is similar to the Put option.

Given that $\sigma, r$ and $T$ are constant within the context of the loop, the implementation
of equation~\ref{eqn:montecarlo}
creates a compute bound process dominated by the exponential function, with 
relatively few loads and stores to memory. 
One bottleneck is the control of numerical round-off error associated 
with a floating point summation process over a large 
number of terms~\cite{Linz:1970,Higham93theaccuracy}. Another  
is the generation of sets of good quality pseudo-random 
numbers (PRN) because the computed price converges only slowly as
$O \left ( \frac{1}{\sqrt{N}} \right )$.
This is significant because the operation count during execution of the 
for-loop code scales as $O(N)$. In our work, uniform random numbers are generated using 
the 32-bit version of the Mersenne Twister algorithm~\cite{mersenne:32}, then 
transformed to a standard normal distribution using the Box-Muller formula. 

\subsection{Binomial Tree (BT)}

The binomial tree option pricing model~\cite{coxrossrubenstein:1979} builds a lattice of options prices with a root node at 
the starting date and multiple nodes at the end date. 
The model decomposes the time variable into discrete steps, each corresponding to a level in the lattice. 
The price can be computed at any level in the tree, which corresponds to an 
intermediate date within the contract, thereby making the binomial 
model suitable for pricing American and Bermudan options. 
In general at each time point $i$, there is a vector of option prices computed, $S_{ij}$.
Given price $S_{ij}$ and a pair of factors $u$ and $d$ representing the up and down movements, 
the two possible prices at the next level are $S_{(i+1)j}=uS_{ij}$ and $S_{(i+1)(j+1)}=dS_{ij}$
where the factors $u,d$ are constant across the tree and 
are formally defined as:

\begin{equation}
   u = {\rm e}^{\sigma \sqrt{T}} \;\;\;\ {\rm and} \;\;\;\; d = \frac{1}{u}
\end{equation}

The complete algorithm therefore has three steps:
\begin{itemize}
  \item Given $S_0$, the current spot price, work forward from today, the date of computation, to the 
        expiration date (timepoint N), applying the up and down factors at each step and thereby
        computing all final node prices $S_{Nj}$.
  \item At each final node of the tree (level $n$) compute the exercise value. 
  \item Iterate backwards from the final nodes in the tree and for every intermediate 
        node compute the option price assuming risk neutrality, which 
        means that the price is computed as the discounted value of the future 
        payoff.
\end{itemize}

The final step; a nested for-loop, dominates the time taken by the computation. The 
number of updates performed scales as $O(\frac{N^2}{2})$ with each update consisting of
two floating point multiplications and one floating point addition. This contrasts
with the MC algorithm, where there is a need to repeatedly compute transcendental 
mathematical functions. 
The convergence of the method is a function of the number of timesteps chosen for the 
computations. 

It is not necessary to hold all of the prices $S_{ij}$ in memory at the same time. Only
the prices at timepoint $i$ are needed. A vector with the 
elements being overwritten as the values are processed during the backwards iteration
process suffices for this purpose. The BT implementation is thus dominated by pointer 
dereferencing to generate array indices and by data move activity, in contrast to the 
random number selection and sum reduction characteristics of the MC algorithm.

\section{Optimization Methodology}
\label{sec:Optimization Methodology}
In this section we discuss a common set of optimizations applied to our workload 
to tune performance. These include algorithmic optimizations and platform-dependent optimizations 
through unrolling and vectorization, either by the compiler or via explicit use of vector instructions. 
We develop a common, platform-independent methodology for these optimizations and discuss
any platform-dependent implementation needs, where necessary.

We developed the initial code base in the C language to be used on all servers
for the experiments. The heart of our system is an OptionPricer
implementing the algorithms described in Section~\ref{sec:option_pricing}.
We manually unrolled the loops and used the same vectorized algorithm on each server, albeit
with modifications to accommodate different widths of the vector ISA supported on each server. 
The common approach to vectorization on all platforms, using either the compiler or a direct algorithmic
translation to vector code, reflects similar optimization effort on each server and 
is the fairest way to perform a comparison among servers. 

\subsection{Algorithmic optimization for Monte Carlo}
In its usual form, the Monte Carlo kernel is a for-loop over a set of normally distributed random 
numbers, $x_i$. The loop applies a formula to each number, followed by a summation. 
Within the loop, each random number is used to compute a payoff value, 
a process which involves the relatively expensive evaluation of an 
exponential function. If this value is positive then it contributes to the 
sum otherwise it is discarded. This means that an if-statement 
dominates the loop. This algorithmic feature hampers further  
performance and power efficiency optimizations, as will be amply clear in the results section.  
Given that all variables in the payoff expression, except the random number, are loop-invariant, a 
threshold, ${\rm Thres}$ for pre-screening the random numbers can be 
applied. For Call options this is defined as:   
\begin{equation}
   {\rm Thres} = 
        \frac{1}{\sigma \; \sqrt(T)} \; {\rm \log}_{\rm e}
        \frac{\rm strike \; price}{ {\rm stockPx} {\rm e}^{ ( r - \frac{\sigma^2 T}{2} ) } }
   \label{eqn:mc_randnos_threhold}
\end{equation}

such that the payoff value is computed using only values $x_i$ where:

\begin{equation}
   x_i > {\rm Thres} 
\end{equation} 

The impact of this screening on any given option contract evaluation 
depends on the specific values of stock and strike prices, interest 
rate, volatility and time to expiry. 

On one hand, screening random numbers as above leads to a further algebraic 
simplification allowing several multiply operations to be factored 
outside the sum loop, which now becomes:

\begin{equation}
   \sum_{j=1}^{M} {\rm e}^{\sigma\sqrt(T) \; x_j} \label{eqn:mc_exp_sum}
\end{equation}

where $M$ is the number of random numbers passing the threshold test defined in 
Equation~\ref{eqn:mc_randnos_threhold}. On the other hand, this prevents complete 
vectorization of random number generation within the compute intensive part, which is also 
dependent on the input data. 

\subsection{Vectorization}

Server processors have diverse vector ISAs and vector unit implementations. In this section 
we discuss whether and how this diversity can be overcome with a common approach for each
of our two financial analytics kernels.

\subsubsection{Monte Carlo}

We leverage the Cephes software library~\cite{iMOS00a} in our vectorization exercise.
The library is an open source collection of high quality mathematical routines written in 
C. These routines have been tested for over thirty years in scientific and 
engineering applications. The library presents a common reference, open source 
alternative to vendor specific libraries and to the math library 
supplied by each C compiler. It implements versions of the $sin$, $cos$, 
$exp$ and $log$ functions tailored for different vector units including 
Intel's SSE\footnote{http://gruntthepeon.free.fr/ssemath/} and 
AVX\footnote{http://software-lisc.fbk.eu/avx\_mathfun/} as well as 
ARM's NEON\footnote{http://www.arm.com/products/processors/technologies/neon.php}. 

We outline the Monte Carlo kernel vectorization 
using the Intel SSE instruction set as an example. The SSE instructions operate 
on 128-bit wide registers, named XMM registers, while the exponential function 
uses single-precision floating point variables, each 32-bits wide. This 
means that four separate values are computed in parallel in a 
single SSE $exp$ instruction using one XMM register. The sum loop, shown 
in Equation~\ref{eqn:mc_exp_sum}, is transformed to perform one fourth 
of the iterations due to the vectorized exponential calculation. Note 
the ARM NEON registers are 128-bit wide too and apart from calling the 
NEON-specialized exponential function, code changes are identical to 
Intel's SSE. For the AVX implementation, YMM registers extend XMM 
registers to be 256 bits wide, thus it is possible to compute eight 
values in parallel. The loop is further transformed to accommodate this 
extra opportunity. 
A further optimization that we have not implemented here, would be to 
extend the vectorized exponential routine to use several more vector 
registers in a single invocation. This could further improve 
performance by reducing the number of iterations in the sum loop and 
the associated overhead. It could also provide additional gains by 
leveraging better locality on cached data. 

\subsubsection{Binomial Tree}

The binomial tree computation is dominated by an innermost loop which 
processes the set of all values at each timestep to generate values 
for a previous timestep. A loop step can be expressed simply as 

\begin{equation}
   x_i = a x_i + b x_{i+1} \label{eqn:bt_loop}
\end{equation}

This loop step relation shows that there is an anti-dependency between 
iterations. This dependency needs to be removed in order to 
enable vectorization. Notably, the GCC compiler 
failed to recognize this vectorizing opportunity on all our platforms, because of this 
dependency. The Intel compiler (ICC) was able to overcome this barrier and 
produce vectorized code on the Phi and Sandy Bridge servers. 

\begin{figure}[htbp]
  \centering
  \includegraphics[width=0.2\textwidth]{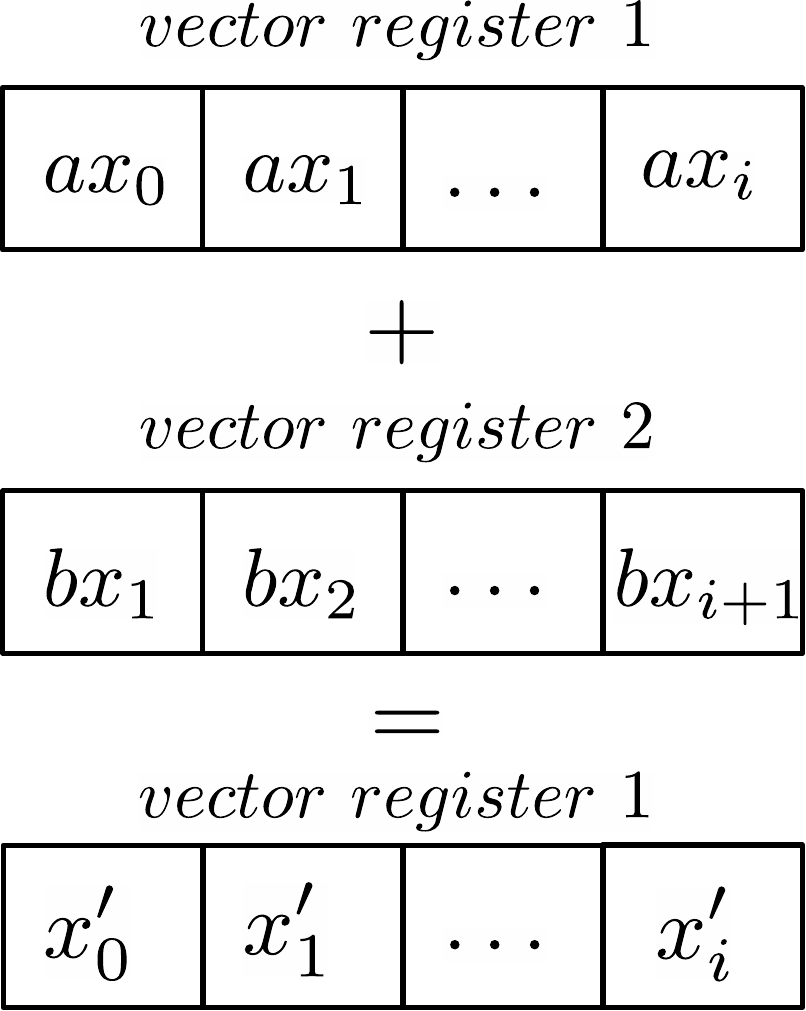}
\caption{Vectorization of Binomial Tree kernel}
\label{fig:vecops}
\end{figure}

We have constructed two handcrafted vectorized solutions for the kernel: 
one which replaces the inner loop entirely by invoking a routine 
written in vectorized assembly, and another using compiler vector intrinsics 
placed inline with the C source code. The latter are referred to as INTRINSICS\_xxx build and runs, 
where xxx depends on the platform. 
These two handcrafted vectorization solutions overcome the apparent anti-dependency 
and enable us to compare the 
low-level manual assembly solution versus a more abstract, intrinsic-based 
vectorization method. Both versions employ loop unrolling  in conjunction with vectorization 
to calculate more values in a single iteration. Figure~\ref{fig:vecops} shows 
schematically the vectorized BT calculation. The unroll factor is equal to the vector register
width measured in double-precision floating point numbers (single-precision in the case of the
ARM Cortex A9 NEON unit).  
Similarly to the Monte Carlo algorithm, we present a detailed example of 
vectorizing the BT calculation, using the the SSE 128-bit instruction set 
as an example. The sequence of vectorization steps is as follows:

\begin{itemize}
  \item Preload high and low double words of xmm10 with value $a$, and xmm11 with $b$.
  \item Populate vector registers xmm0,1,2,3 with eight double precision (64-bit)
        contiguous values, two per register.
  \item Copy xmm0,1,2,3 to xmm5,6,7,8 and downshift the eight values, using register
        shift and OR operations.
  \item Load one value to fill the gap at xmm8 caused by the downshift.
  \item Execute the four products and summations given symbolically by 
        \begin{gather*}
           xmm0 = xmm0 \cdot xmm10 + xmm5 \cdot xmm11 \\
           xmm1 = xmm1 \cdot xmm10 + xmm6 \cdot xmm11 \\
           xmm2 = xmm2 \cdot xmm10 + xmm7 \cdot xmm11 \\
           xmm3 = xmm3 \cdot xmm10 + xmm8 \cdot xmm11 \\
        \end{gather*}
  \item Write the registers xmm0,1,2,3 to RAM
  \item Repeat the steps above for as many sets of eight values
        as exist in the loop.
\end{itemize}

We follow a similar vectorization approach for the ARM NEON 128-bit 
vector ISA, the Intel AVX 256-bit vector ISA and the 
Intel Xeon Phi Knight's Corner (KNC) 512-bit vector ISA. 
A difference between the Intel family and ARM NEON vector units is 
that NEON supports only single-precision floating point vector operations. 
Therefore, the vectorized calculation of Binomial Tree on ARM has double the throughput of 
SSE and effectively the same throughput as AVX.
While in our experiments we execute enough iterations of the financial analytics kernels to 
guarantee production-strength numerical accuracy, studying in depth the implications of numerical 
imprecision due to the use of 32-bit vector floating point arithmetic is beyond the scope of this paper.

As a further optimization, we attempted to reduce memory loads as much as 
possible by initially loading data values to registers and then manipulating 
these registers to build the operands for the Binomial tree calculation. ISA 
support for this manipulation is an important differentiating point between 
the vector units of our server processors.
SSE provides inter-lane shift (\texttt{psrldq}, \texttt{psllq}) operations on XMM registers and building 
the second operand of the Binomial Tree kernel (vector register 2 in Figure~\ref{fig:vecops})
is possible with three vector instructions. However, the AVX instruction set 
lacks inter-lane shifts. This results in an intricate sequence 
of permutation and blend operations (\texttt{vpermilpd}, \texttt{vperm2f128}, 
\texttt{vblendpd}) needing seven instructions.  
Notably, the NEON (\texttt{vext.32}) 
and KNC (\texttt{valignd}) instruction sets provide a vector align (also known as 
vector extract) operation. This operation combines elements individually from 
different vector registers and allows building the Binomial Tree calculation operand with 
a single instruction. 

Moreover, of all the vector instruction sets, only KNC provides a 
\emph{fused multiply-add} operation (\texttt{vfmadd132pd}). This means
that with the KNC ISA, the calculation shown in Equation~\ref{eqn:bt_loop}, 
requires only two vector instructions while with the other ISAs 
the same calculation needs three instructions to do the vector multiplications and 
the final summation. 

\subsection{Compiler Based Vectorization}

To test compiler-based vectorization (labeled AUTOVECT in our analysis) with GCC, we first used the same GCC compiler flags on all servers: \mbox{-O3 -Wall -g -fPIC}.
We linked the libevent (needed by the data feed handler, to be discussed further in Section~\ref{sec:Experimental Setup and Measurement Methodology}), ipmi-1.4.4, math and Pthread libraries statically during the compilation process. 
To test vectorization with ICC on the Intel Sandy Bridge server, we used the ICC -mavx -ftree-vectorizer-verbose=0 flags.
For the Intel Xeon Phi server we used the ICC -mmic -vec-report=1 flags.
As for ARM,  we used the -mcpu=cortex-a9 -mfpu=neon -ftree-vectorizer-verbose=0 GCC compiler flags.
Alternatively the vectorization flags were all disabled for the plain vanilla no-vectorization (labeled NOVECT) builds and runs.

The metrics that we report across all servers reflect a fixed development effort of approximately 30 days
to create and test the code base.

\section{Experimental Setup and Measurement Methodology}   
\label{sec:Experimental Setup and Measurement Methodology}

Our experimental setup includes three platforms on which we execute the
OptionPricer and collect workload-specific performance and energy metrics. 
The next Section defines the metrics that we report, Section~\ref{sec:platforms:platforms} 
describes the platforms and Section~\ref{sec:platforms:methodology}
provides details of the methodology used to obtain the power readings and calculate the energy consumption.  

\subsection{Definition of Metrics} \label{sec:platforms:metrics}

Option pricing in finance takes place by consuming 
a live streaming data feed of stock market prices, often within the context of high 
frequency trading (HFT) for the purpose of pre-trade risk analytics.
The execution time characteristics of option pricing are different 
from those of numerical simulation in computational science using HPC. One aspect is that
scientific codes tend to have measurable setup and post-processing phases, often 
dominated by integer arithmetic, while the main computational phases
are dominated by floating point computations in loops. These phases tend to have
radically different performance and power profiles. By contrast, option pricing runs 
relatively small standalone kernels at very high frequency with little set up 
and post processing phases.
Option pricing on live market data is thus more similar to distributed event processing. Motivated
by the distinctive features of real-time option pricing, we present and use three workload-specific metrics to compare servers under financial analytics workloads:

\begin{description}
	\item[Joules/option] (J/Opt)
         The energy consumed per execution of a pricing kernel is
         a fundamental metric given that this step is repeated at very 
         high frequency throughout a trading session. In the case of 
         an actively traded stock, with a high 
         number of defined option contracts, this building
         block is repeated throughout the trading day without a break.
         Correspondingly, a reduction in this value can translate to significant
         cost savings for providers offering  
         option pricing services within their portfolio.
	\item[Time/option] (S/Opt)
         In contrast to providers, end users, particularly those engaged in HFT, 
         are sensitive to end-to-end latency, thereby constraining the elapsed 
         time per option metric, which in turn defines the total time to price all
         contracts for a given stock. Option pricing has this 
         time-to-solution performance metric in common with HPC applications. 
	\item[QoS]
		New prices may arrive at any time in a trading session. This means that 
		any contracts not yet priced using the previous price update are 
		abandoned and deemed unusable. Related to the Time/option metric, 
		but also dependent on market activity, we define the Quality of Service 
		metric (QoS) as the ratio of successful to the total requested option 
		pricings. The QoS metric is an application-specific measure on meeting 
		option pricing performance requirements. It is useful for 
		characterizing application-related performance and scalability offered 
		by deploying multiple nodes. It is worth noting that QoS depends on the 
		stock price change rate and other market activities at the time of its calculation.
		This implies that QoS will vary each time each time it is calculated in a live market scenario.
\end{description}
  
\subsection{Hardware Platforms}  \label{sec:platforms:platforms}

We used three platforms, one state-of-the-art server architecture with Intel Sandy Bridge 
processors --briefly referred to as ``Sandy Bridge'' in the rest of this paper, one state-of-the-art HPC 
server with Intel Xeon Phi Knights Corner coprocessor --referred to as ``Xeon Phi''
 and a Calxeda ECX-1000 microserver with ARM Cortex A9 processors, packaged in a Boston Viridis
rack-mounted server --referred to as ``Viridis''. 
We used the $4.7.3$ version of the GCC compiler and the Intel Compiler ICC version $14.0.0 20130728$ 
for code generation.  The three platforms offer the possibility of 
scaling their frequency and voltage through a DVFS interface.
We conducted experiments with the highest and lowest voltage-frequency settings on each platform,
to which we refer as performance mode experiments and powersave mode experiments respectively.

Other details of the two platforms are as follows:
\begin{description}
	\item[Sandy Bridge] is an x86-64 server with two Intel Xeon CPU E5-2650 processors 
	operating at a default frequency of 2.00GHz and equipped with 8 cores each. 
	The machine has 32GB of DRAM (4 $\times$ 8GB DDR3 @ 1600Mhz).
  The frequency in powersave mode is $1.2$ GHz and $2.0$ GHz in performance mode. 
	The server runs on Linux CentOS 6.5 with kernel version $2.6.32$ ($2.6.32-431.17.1.el6.x86\_64$).
   
	\item[Xeon Phi] (Knights Corner) is a 5110P co-processor board over PCIe. It is based on the many 
	integrated cores (MIC) architecture running at a default frequency of 842.104MHz. It targets the high 
	performance computing (HPC) market a highly parallel design and energy efficiency.  High performance is 
	the result of using many cores, dedicated vector processing units with wide vector registers, hardware 
	transcendentals, L2 hardware prefetching and 6 or more GB of GDDR5 as a DRAM cache on-board.  High energy 
	efficiency is achieved through the use of a low clock frequency and simple x86 cores with lightweight 
	design, both suitable for parallel HPC applications. KNC supports 64-bit instructions and provides 512-bit 
	SIMD vector registers. In performance mode the frequency is 1.053GHz and in powersave mode it is 842.104MHz.
	The Phi has 60 cores and each core is capable of 4-way hyper-threading. 
	The system runs Linux kernel 2.6.38.8+mpss3.2.1.
  
	\item[Viridis] is a 2U rack mounted server containing sixteen  
	microserver nodes connected internally by a high-speed 10 Gbps Ethernet network. 
	This means that the platform appears logically as sixteen servers within one box. 
	Each node is a Calxeda EnergyCore ECX-1000 comprising 4 ARM Cortex-A9 cores and 4~GB of DRAM
	running Ubuntu 12.04 LTS. 
	Viridis has a frequency of $1.4$GHz in performance mode and  
	a frequency of $200$MHz in powersave mode.
\end{description}   

When referring to the different platform settings we will be using the following notation
 to represent the platform configuration [Nodes used $\times$ Cores Used $\times$ Threads per Core].
\phantom{force a blank line manually here.} 

\subsection{Methodology}  \label{sec:platforms:methodology}

In this section we provide detail of our proposed experimental and measurements methodology, which are
both critical for fair comparison between our  platforms.
\subsubsection{Experiment Methodology}  \label{sec:platforms:experiments}
\begin{figure}[htbp]
  \centering
  \includegraphics[width=0.4\textwidth]{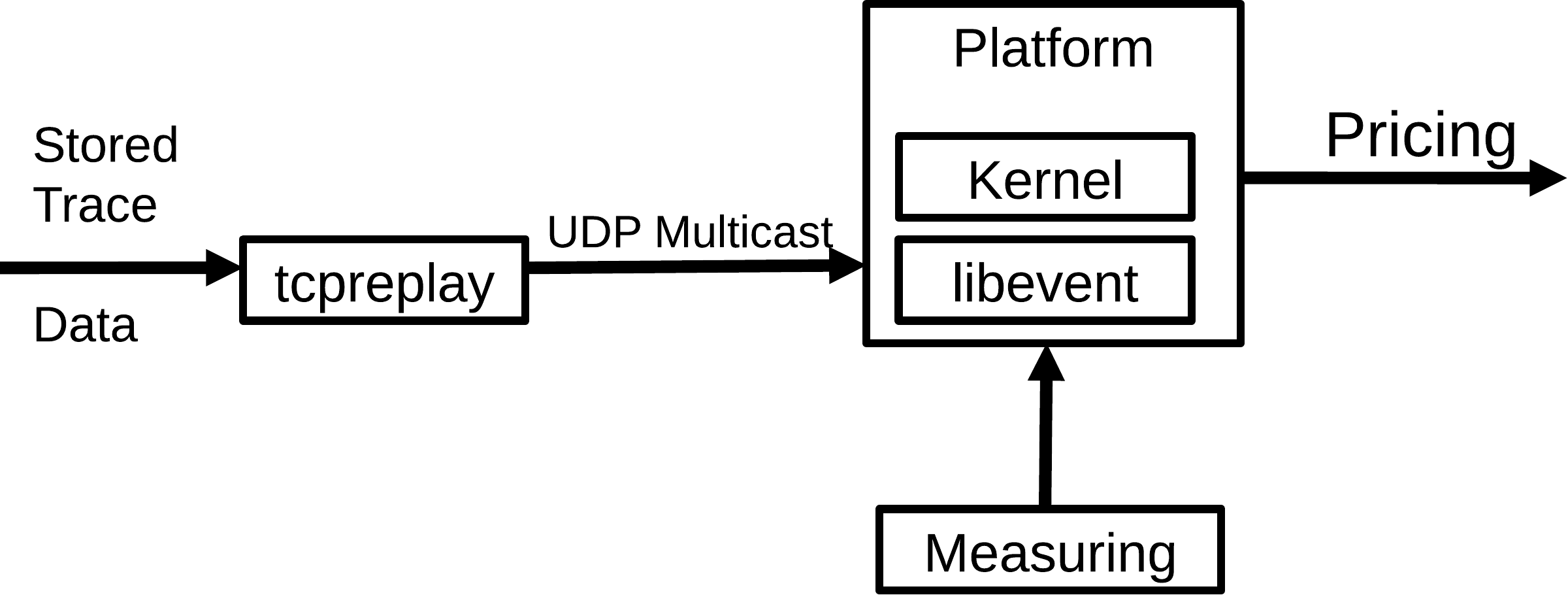}
  \caption{Financial trace data measurement setup}
 \label{fig:fig1}
\end{figure}

We collected Facebook stock price ticks during a full New York Stock Exchange session and replayed them using UDP multicast to all nodes in each platform, as shown in Figure~\ref{fig:fig1}. This is as close as an experiment needs to be to reality without any external factors affecting the setup or measurements. 
We used \texttt{libevent} to capture the event of an option price changing, then the OptionPricer to 
calculate new prices for 617 Facebook options at the maximum speed feasible.
It is worth noting that \texttt{libevent} is only used for its capacity to trigger option pricing
events throughout this work.

\subsubsection{Power and Energy Measurements}  \label{sec:platforms:measurements}

\begin{figure}[htbp]
\centering
  \includegraphics[width=0.45\textwidth]{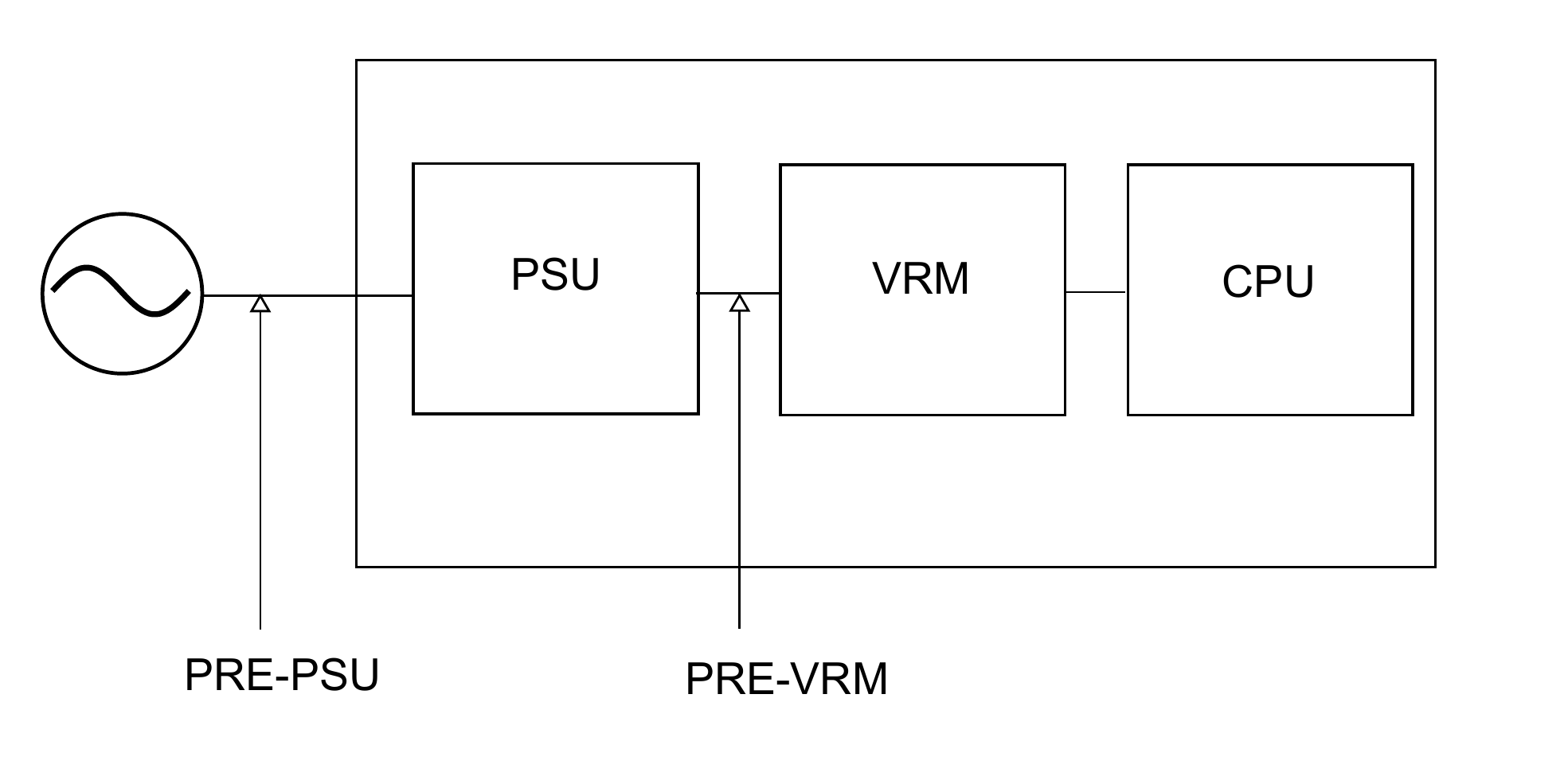}
  \caption{The path of the current supply to the CPU showing points at which
           we measured power.}
  \label{fig:currentsupplypath}
\end{figure}

Power measurements can be taken at two points along the path supplying 
current to the CPUs, as shown in Figure~\ref{fig:currentsupplypath}. 

Each measurement path 
exhibits different characteristics~\cite{nikolopoulos:2012}. The Power supply unit (PSU) converts 
the AC wall socket supply to DC, but can be up to $30\%$ inefficient. The voltage regulator module 
(VRM) stabilizes the DC supply before it reaches the CPU. 
The exact form of the current supply path differs from one platform to the next. To provide 
a fair basis for comparison we identified two distinct points on the path 
which are measurable on both platforms.

We use a WattsupPro multimeter to measure the power at the point before the PSU, which we label
PRE-PSU, giving a value that corresponds to the true economic cost for operating each platform. This would 
give an overall picture of the external energy budget used by the different platforms. However, the energy 
consumed internally by the compute cores is also relevant to our study, as it isolates the energy effects
of processors and discards artifacts of cooling and packaging, the study of which is beyond the scope of 
this work. 

To isolate the energy consumption of processor packages, we capture power consumption at the point before 
the VRM, which we label PRE-VRM.
For the Sandy Bridge server, PRE-VRM measurement is facilitated by reading the Running Average Power 
Limit (RAPL) counters while the same functionality on Viridis is available through
the Intelligent Platform Management Interface (IPMI) counters, which is also available on the Xeon Phi platform. 

\begin{figure}[htbp]
   \centering
   \includegraphics[width=0.48\textwidth]{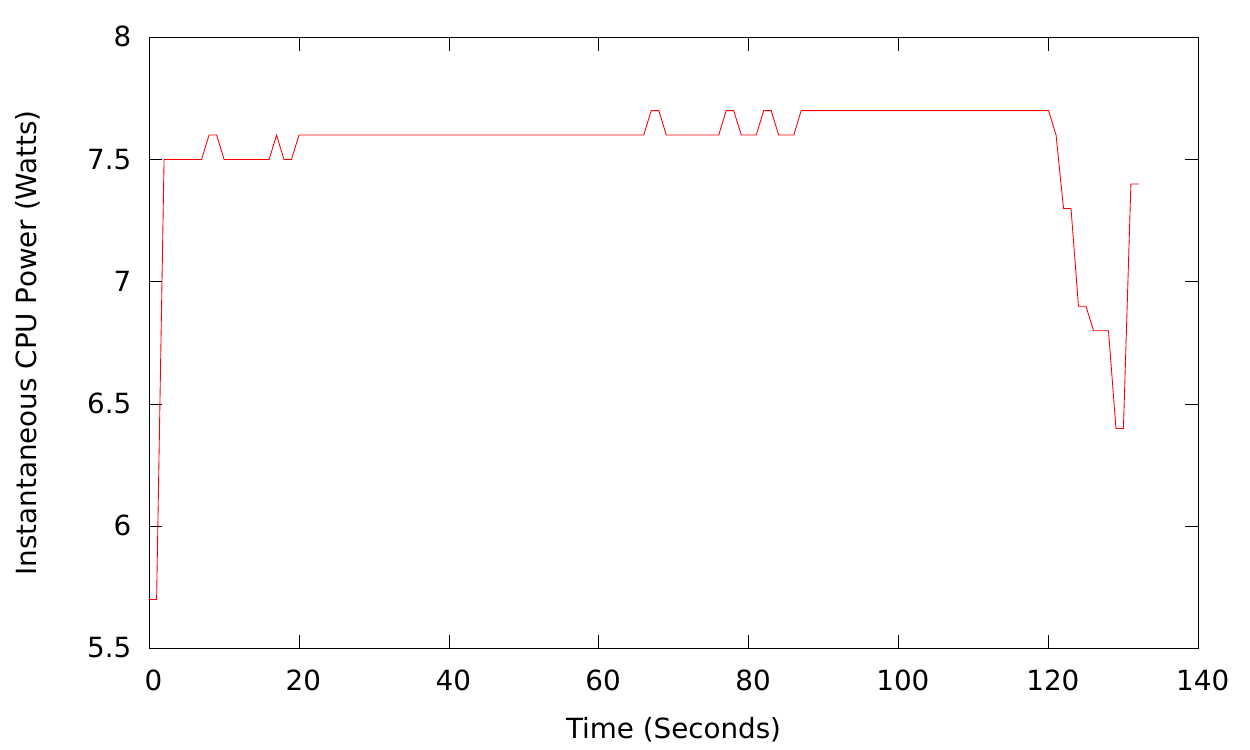}
   \caption{CPU power vs. time for the MC kernel}
   \label{fig:powervstime}
\end{figure}

Figure~\ref{fig:powervstime} shows the power versus time plot for a 
standalone execution of the MC kernel. The BT execution plot is similar. 
The profile of instantaneous power versus time follows 
a very sharp trapezoidal shape: the CPU is fully utilized during execution and 
there are no periods of inactivity. This is a common feature with other 
numerically intensive HPC applications. It means that the measured 
average power is a representative measure of energy consumption 
throughout kernel execution. 

In each experiment with multiple cores or nodes, we measure the worst execution time across all cores 
and use it as the elapsed time for all pricings.

%
%
   %
\section{Experiments and Results}   \label{sec:experimentsandresults}
  
We conducted two sets of experiments, one set where we executed the kernels in standalone mode
and a second set where we executed the kernels in production mode, where each kernel is iterated
to converge to an acceptable solution from a market perspective.
We discuss the standalone kernel experiments and our findings in this section and the production experiments in the Section~\ref{sec:experimentsandresults2}. 

\subsection{Kernel Experiments} \label{sec:experiments_and_results:experiments_performed}   
Stock price data was recorded for a full session of a normal trading day in July 2014
and subsequently used to perform all experiments. By normal we mean that no 
significant initial public offerings (IPOs)  or other skewed trading patterns
took place on the market that day. 
We compiled and executed our OptionPricer on our laboratory 
servers and configured the installation to compute prices using MC and BT models
for the 617 option contracts defined on the Facebook stock at the time.
Each instance of the OptionPricer used as many compute threads as needed to fully 
subscribe the CPUs on all platforms.

Both the MC and the BT models are numerical approximations to the Black-Scholes 
analytical model, converging to this value as the number of iterations or steps $N$ approaches
infinity. We 
selected three values of N for the MC computation 0.5M, 1M, 2M confirming 
that the relative error with respect to the Black-Scholes result decreased
progressively to less than 0.1\%. We found that to achieve comparable
results with the BT model required N equal to 4000, 5000 and 7000 
respectively.
Fixing these values for $N$ in all simulations allows more insightful comparison 
of metrics across models.  

For all experiments reported here, we replayed the collected data onto 
UDP multicast channels in our lab using \texttt{tcpreplay}.  This enabled us 
to conduct the required kernel experiments.
A single price update for one stock is  
pushed onto the multicast channel where it is read by 
each instance of the OptionPricer. All defined
option contracts for that stock are
then computed. Typically there are only a few hundred 
contracts open for one stock, meaning that these experiments
are relatively short to run. These experiments allow precise 
investigation of the J/Option and Time/option metrics.

In our first set of experiments we investigated the effect of performance and 
powersave modes on energy consumption. Tables~\ref{tab:vpowersave}
,~\ref{tab:ipowersave}
and~\ref{tab:xpowersave} (in the Appendix)
show average power, compute time and and energy per option for each 
platform, executing the MC 1M and BT 5000 kernels and measuring the 
PRE-VRM power. Comparing the J/Opt values between performance and 
powersave modes, shows that performance mode is more 
energy efficient on all platforms and in both kernels. This 
is because computation latency under powersave execution increases 
disproportionately to power savings. 

Specifically, on the Intel Sandy Bridge server, 
the PRE-VRM power consumption in powersave mode is about 2/3 of that in 
performance mode one, while energy is consistently worse. By contrast, computation latency doubles,
thus negating any power savings. Note that in powersave mode the Intel CPU runs 
at 1.2 GHz, which is less than half of the performance mode frequency. This
directly translates to more than a twofold increase in latency. 

Considering the Viridis server numbers (Table~\ref{tab:vpowersave}), the increase in J/Option in powersave mode is even more 
pronounced. Powersave power consumption is again a little less than 2/3 of the 
power consumption in performance mode, but computation latency surprisingly increases by an order of 
magnitude. The Viridis powersave frequency (200 MHz) is 1/7 
of the performance mode frequency (1.4 GHz). Apparently this does not 
translate to proportional power savings, because the ARM SoC includes 
several components which are not controllable through the DVFS interface. 

The powersave mode measurements for all configurations of different platforms show a similar trend to different scales, viz. the powersave mode is not as energy efficient as the performance mode. 
Having observed this trend, all the results shown in the rest of this paper 
are in performance mode unless explicitly stated.

\subsection{Other Findings}
\label{sec:findings}
In this section we report on additional findings based on our kernel experiments. Kernels are 
run using different vectorization approaches, described in Section~\ref{sec:Optimization Methodology}. 
To recap, we use the label NOVECT when the kernel code is not vectorized and AUTOVECT when we set the compiler 
flags to perform automatic vectorization. Moreover, we denote manual vectorization by referring to the 
specific instruction set and bit-width for each platform, e.g., NEON128 for the ARM cores. When manual 
vectorization uses compiler intrinsics instead of assembly, we denote this as INTRINSICS\_xxx, e.g., 
INTRINSICS\_NEON128, or as INTR-xxx for brevity.

\begin{figure}[htbp]
\centering
\subfigure[Viridis(1x4x1)]{
  \centering
  \includegraphics[width=0.45\textwidth]{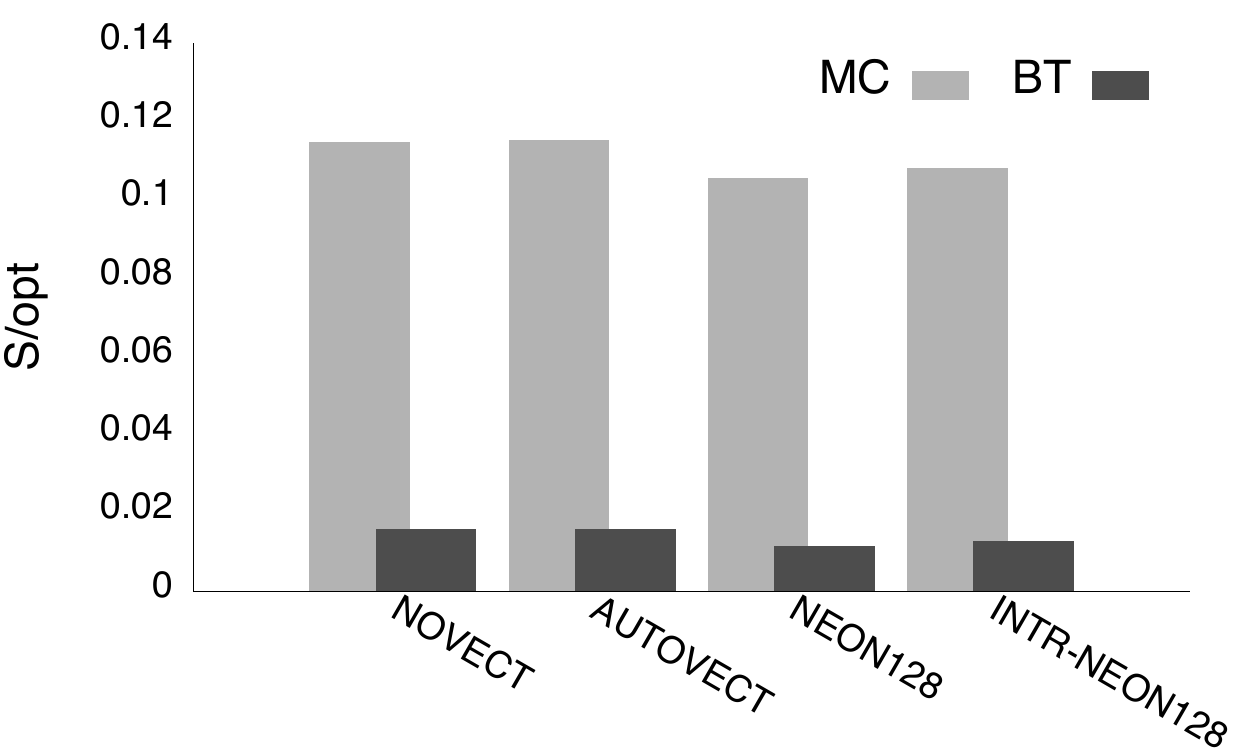}
  \includegraphics[width=0.45\textwidth]{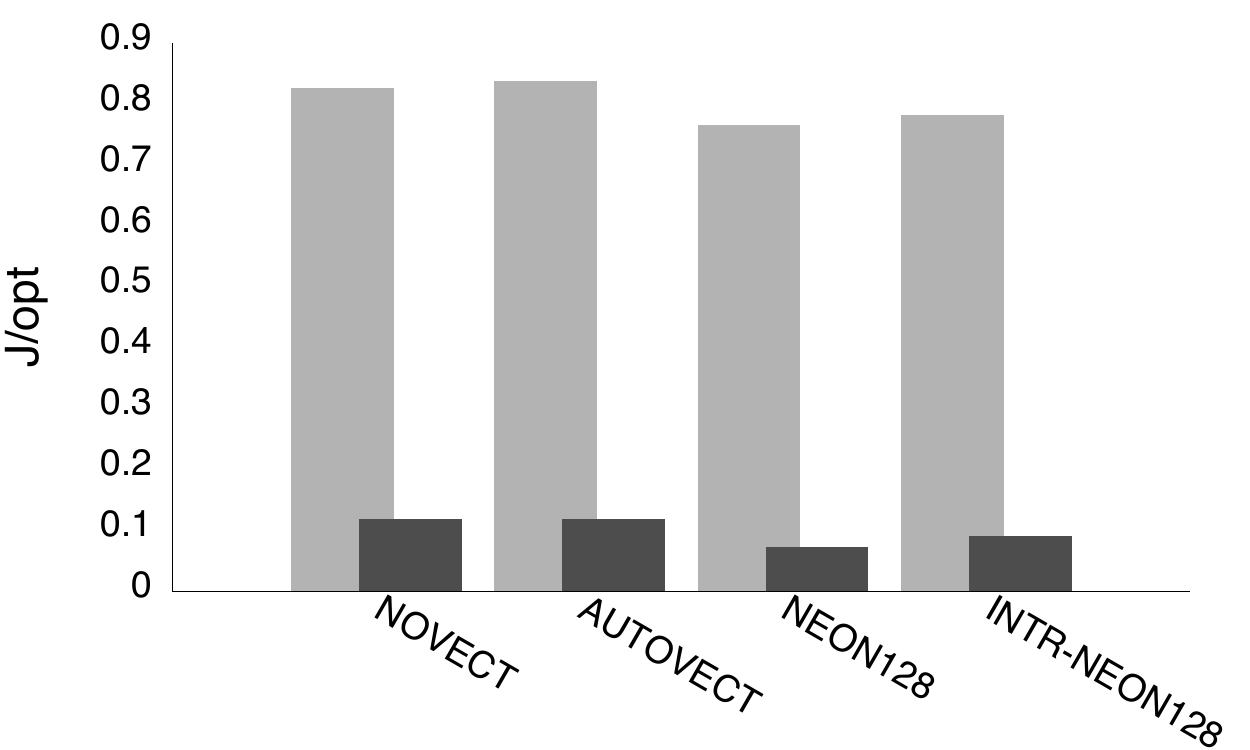}
}
\caption{The cost of non-native transcedental operations on Viridis(1x4x1)}
\label{fig:figz1}
\end{figure}
\subsubsection{Support for transcendental functions}
As discussed in Section~\ref{sec:option_pricing}, the MC kernel depends heavily on the calculation of 
transcendental functions, specifically the exponential. By contrast, the BT kernel spends most of 
the computation on floating point additions and multiplications. Additionally, Viridis ARM cores do 
not support native execution of exponentials while Intel platforms use native hardware for 
transcendental functions. Figure~\ref{fig:figz1} illustrates the effect of transcendental function 
support on S/Opt and J/Opt metrics for Viridis, contrasting the MC and BT kernels. 

One can easily notice this by capturing the difference between MC and BT algorithms on Viridis, which is an order of magnitude, while the same difference is much less on Intel's Sandy Bridge (single node numbers in Tables~\ref{tab:vpowersave},~\ref{tab:ipowersave}) and on Xeon Phi 
(Table~\ref{tab:xpowersave}). 

\subsubsection{Powersave governor mode}
\begin{figure}[htbp]
\centering
\subfigure[MC Sandy Bridge (1x8x1)]{
	\centering
	\includegraphics[width=0.45\textwidth]{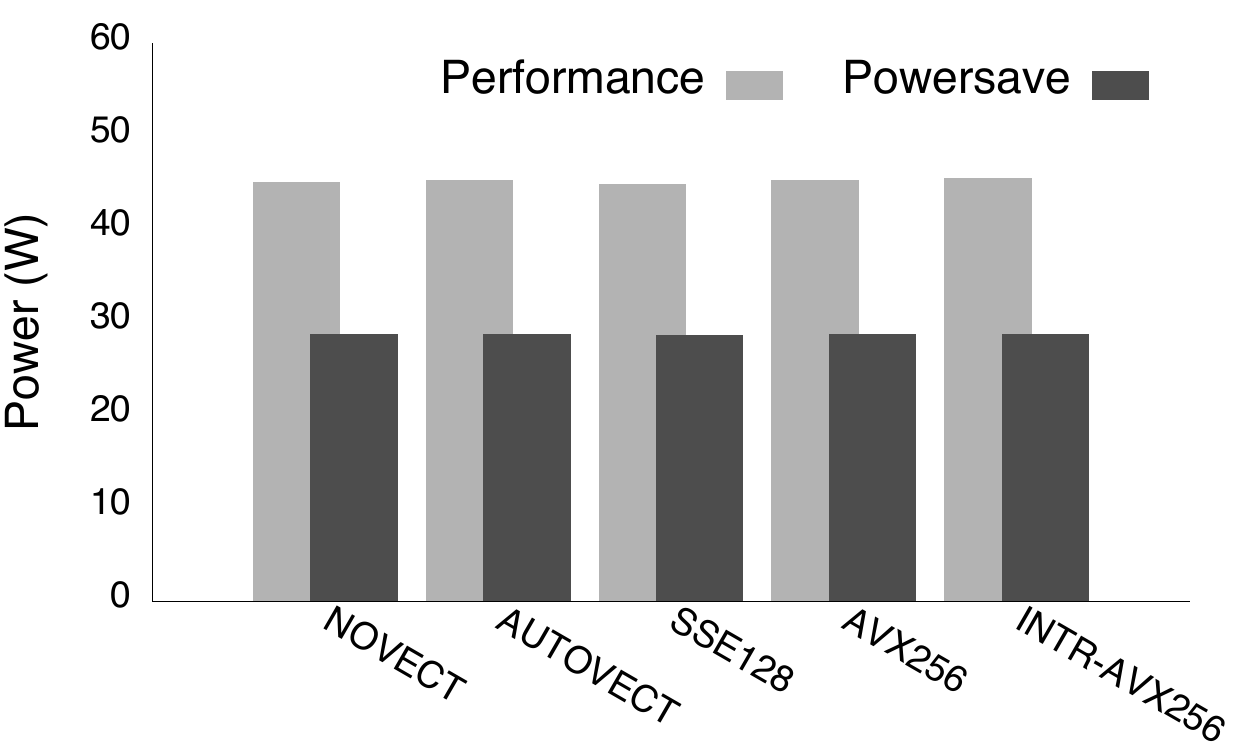}
	\includegraphics[width=0.45\textwidth]{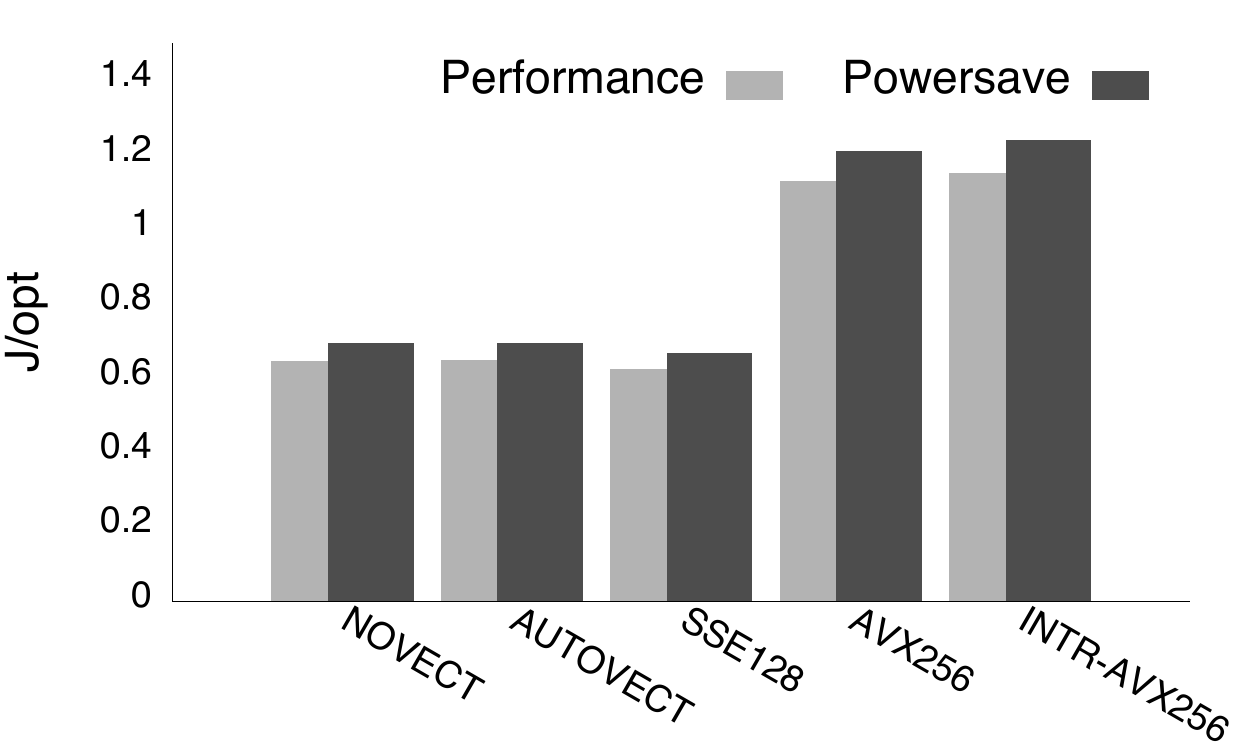}
}
\subfigure[MC Sandy Bridge (2x8x1)]{
	\centering
	\includegraphics[width=0.45\textwidth]{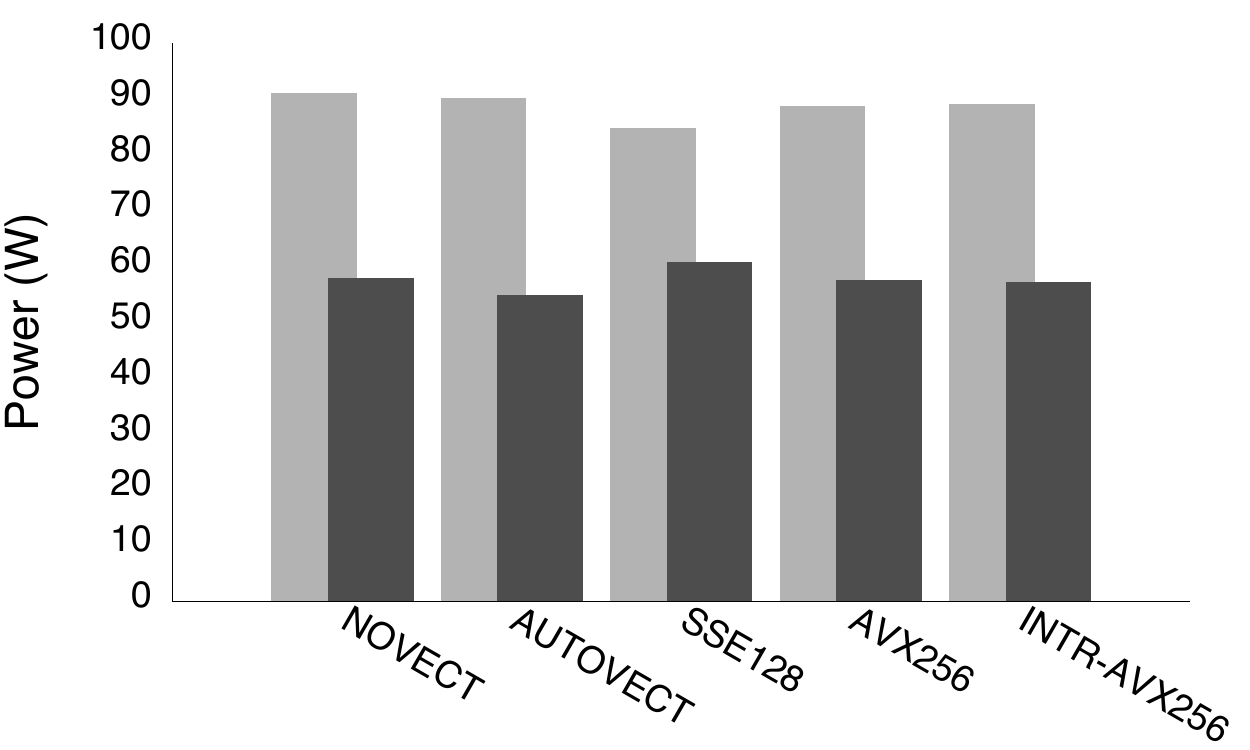}
	\includegraphics[width=0.45\textwidth]{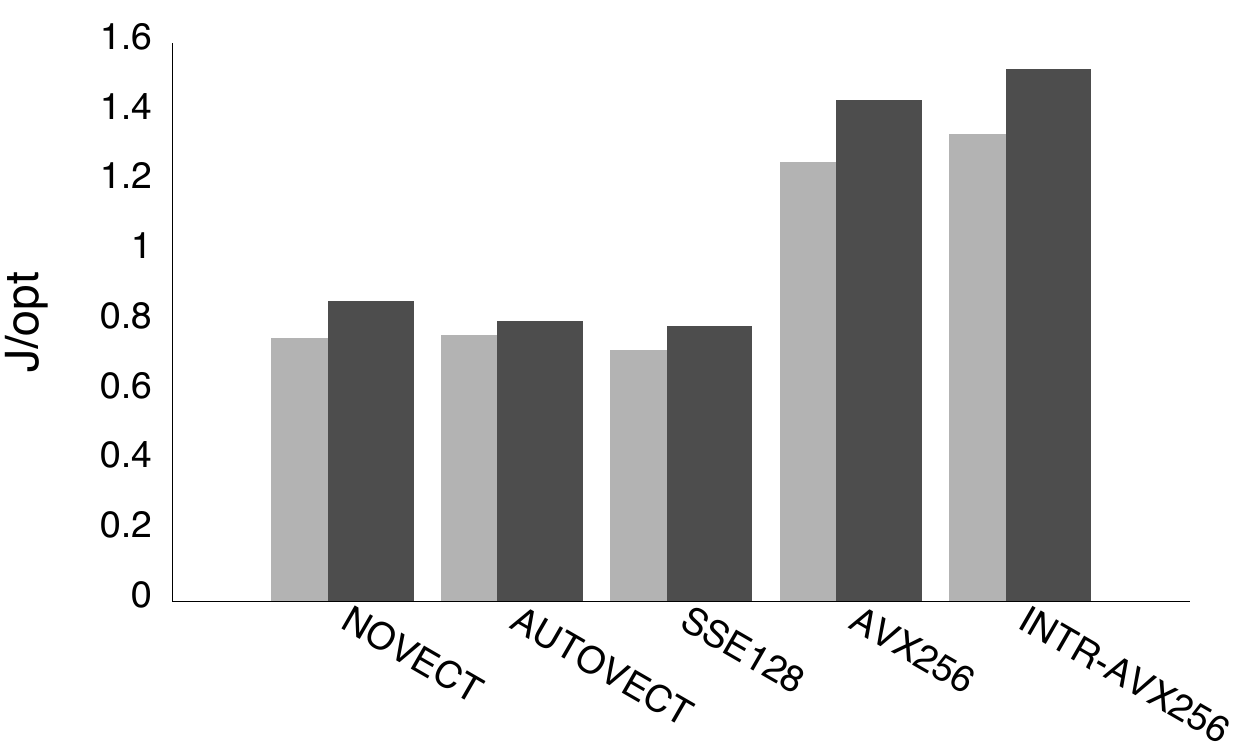}
}
\subfigure[BT Sandy Bridge (1x8x1)]{
	\centering
	\includegraphics[width=0.45\textwidth]{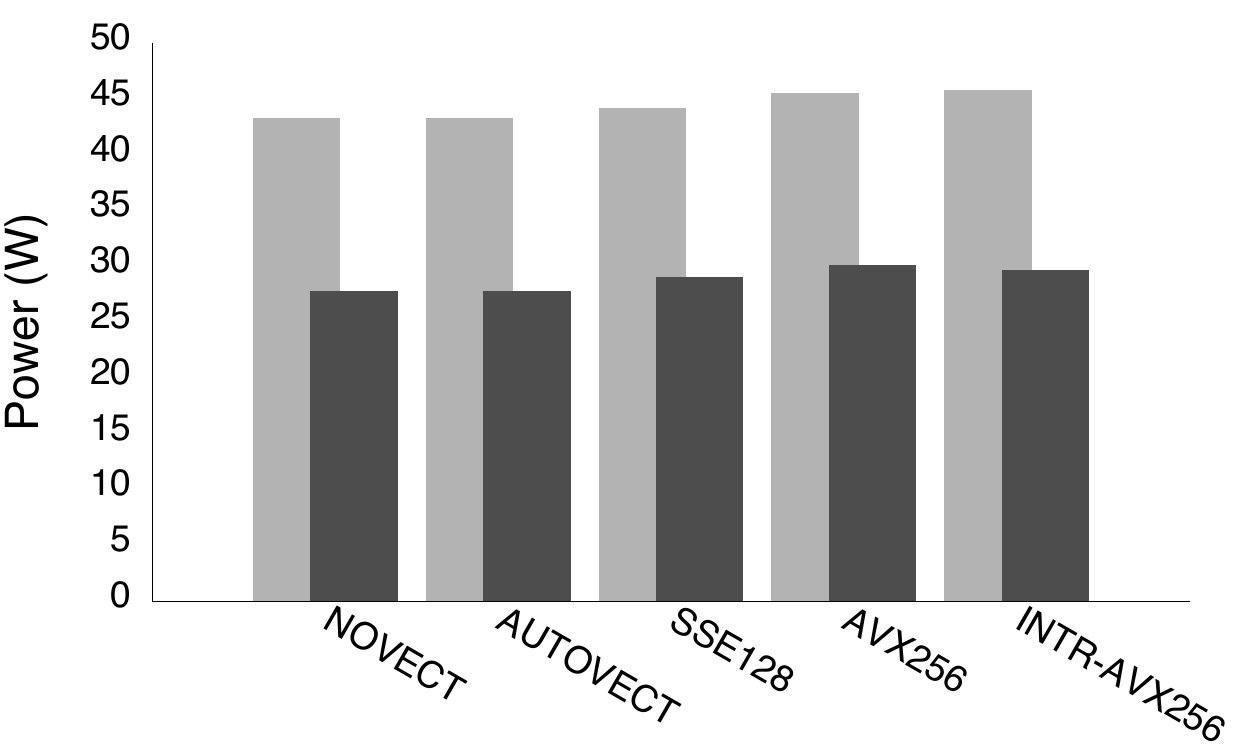}
	\includegraphics[width=0.45\textwidth]{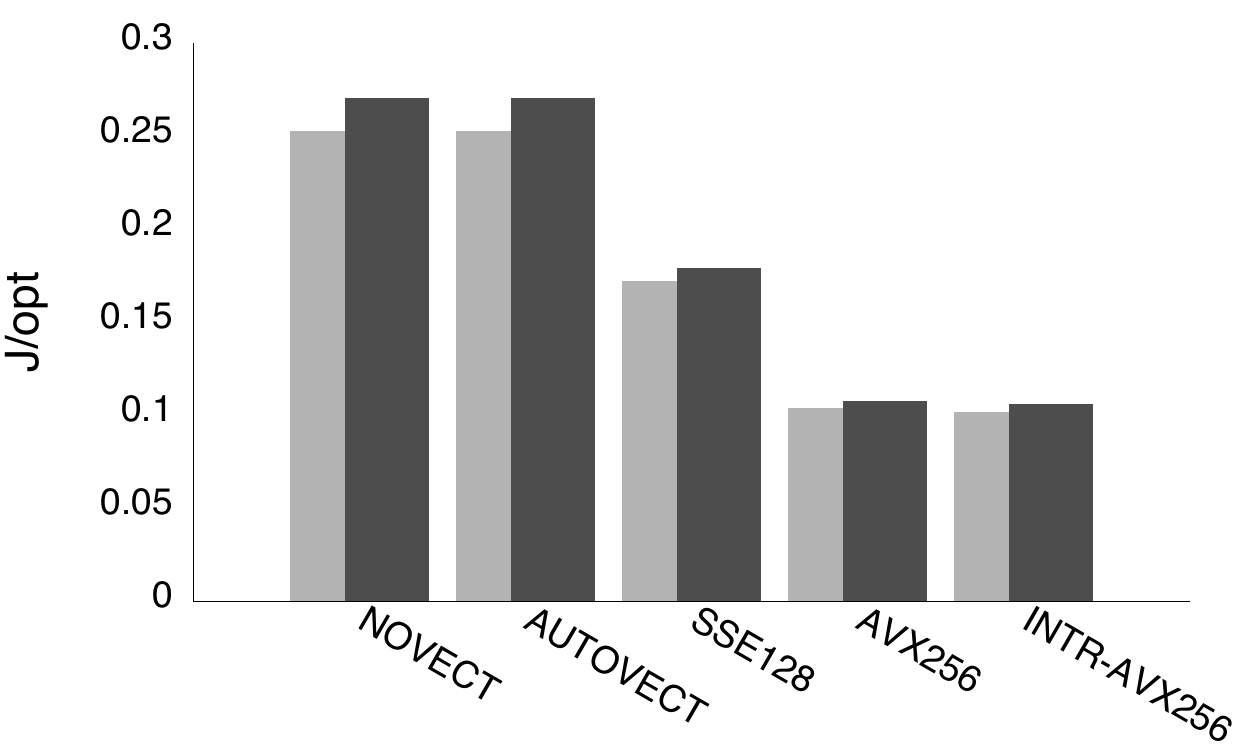}
}
\subfigure[BT Sandy Bridge (2x8x1)]{
	\centering
	\includegraphics[width=0.45\textwidth]{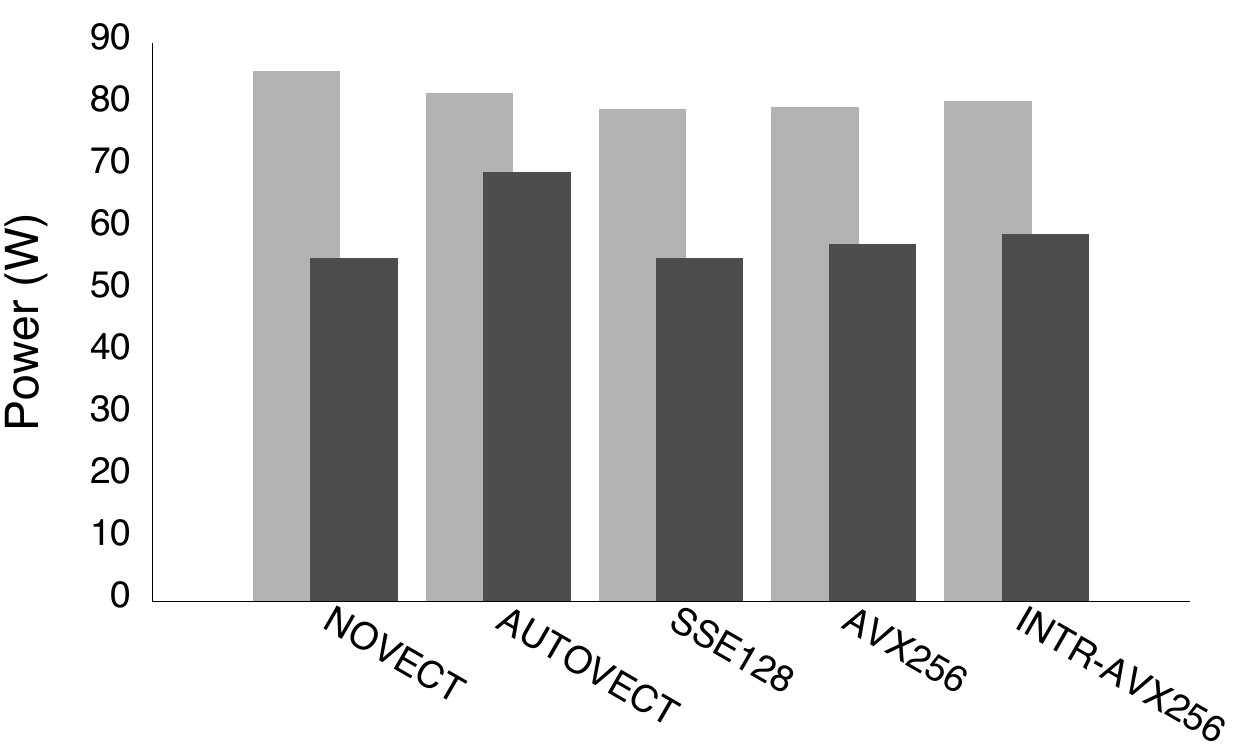}
	\includegraphics[width=0.45\textwidth]{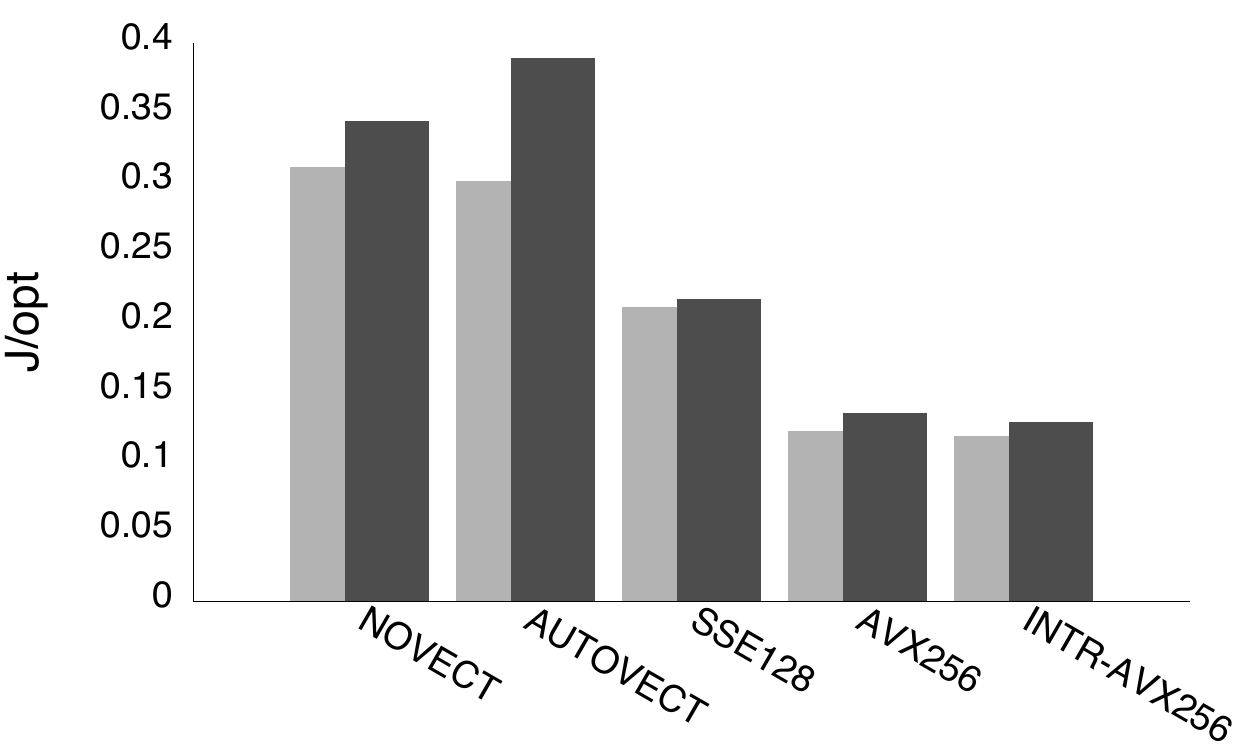}
}
\caption{Power and energy efficiency on Sandy Bridge  platforms for all kernels}
\label{fig:figz2}
\end{figure}

\begin{figure}[htbp]
\centering
\subfigure{
	\centering
	\includegraphics[width=0.45\textwidth]{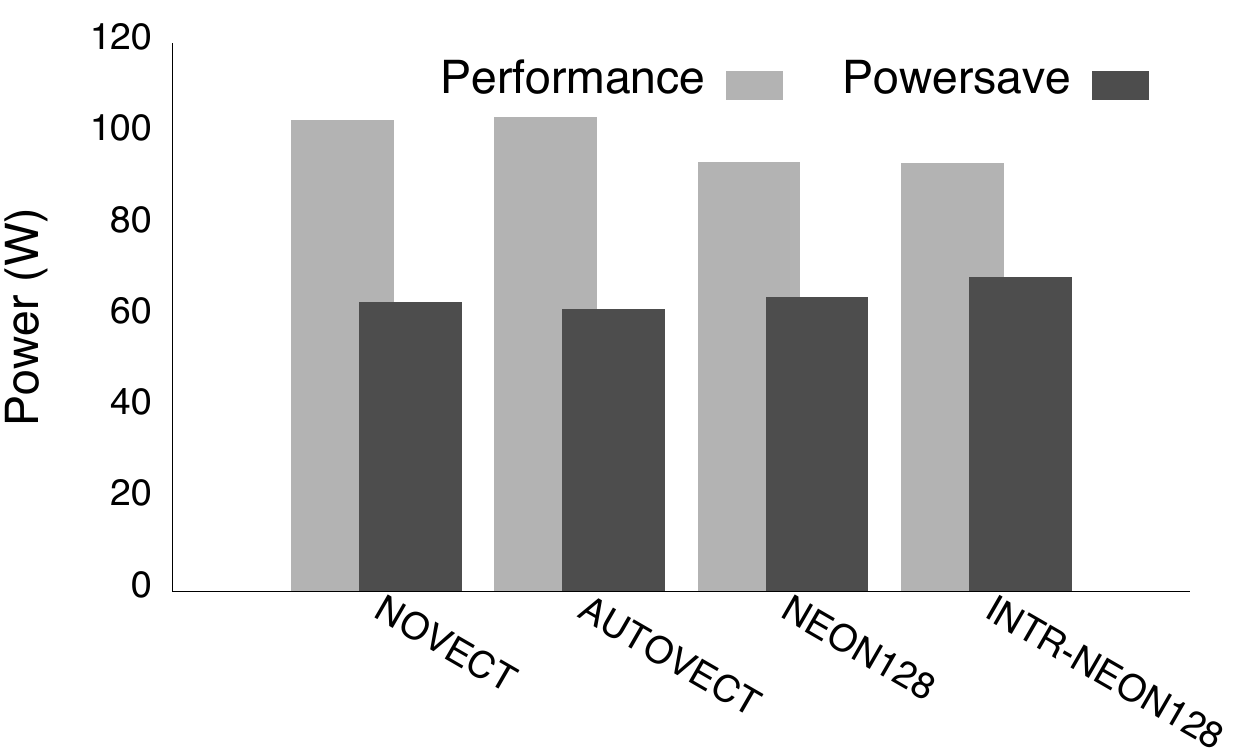}
}
\subfigure{
	\centering
	\includegraphics[width=0.45\textwidth]{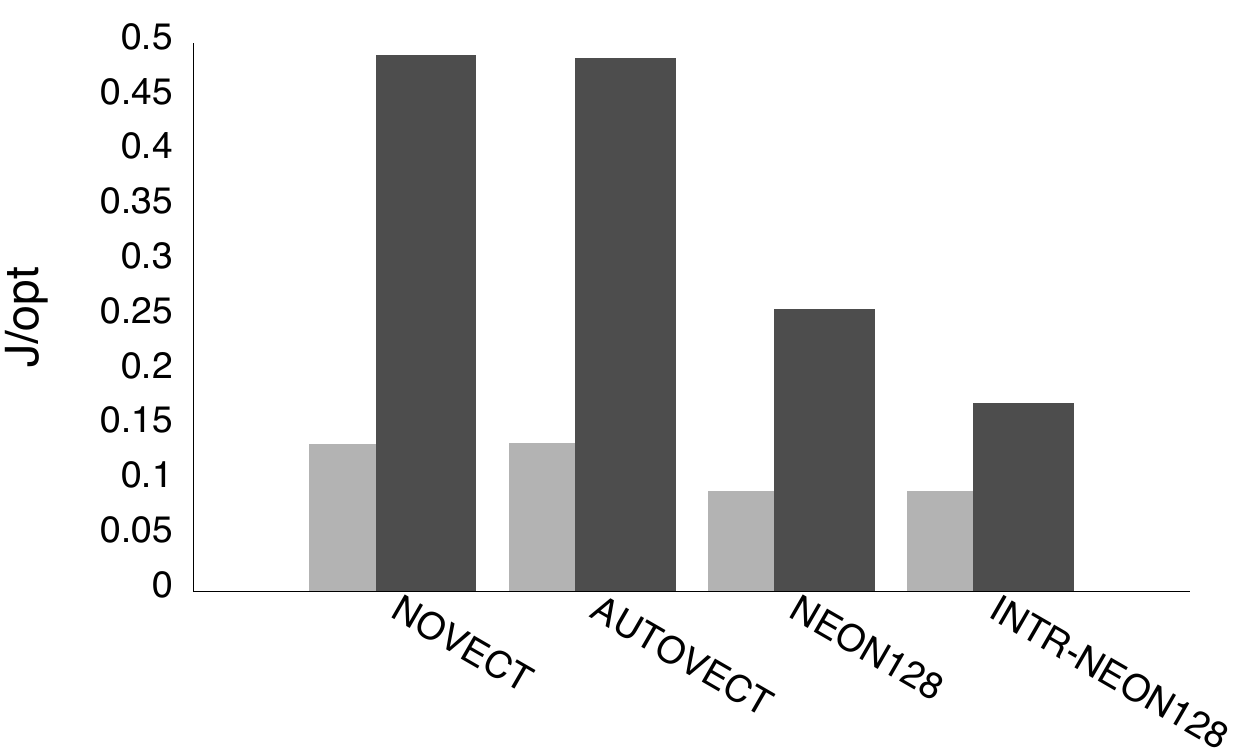}
}
\caption{Power and energy efficiency on Viridis(16x4x1) for the BT kernel}
\label{fig:figz4}
\end{figure}

As previously discussed and mentioned here for completeness, performance mode is more energy efficient than 
powersave mode, despite the latter having a lower voltage-frequency setting. Figures~\ref{fig:figz2} and ~\ref{fig:figz4} 
illustrate this result. Although power consumption is reduced significantly in powersave mode, the energy consumption 
increases for both Viridis and Sandy Bridge, using any vectorization method. Notably, J/Opt for Viridis is up to an order of 
magnitude less in performance mode than powersave mode while on Sandy Bridge the difference between them is much less.
The reduction in power due to powersave mode is lost due to prolonged execution time and results
in similar or higher energy consumption than performance mode. It is worth noting that vectorization in powersave mode recovers some of the lost 
performance, thus the J/Opt difference between the two modes shrinks. 
Comparing the J/Opt numbers for Xeon Phi in powersave and performance modes (Table~\ref{tab:xpowersave}), yields inconclusive 
results on the energy efficiency of the two modes. We are still investigating these results.

\subsubsection{Vectorization Techniques}

The BT and the MC algorithms perform better using intrinsics-based vectorization, than manual vectorization
or autovectorization. This is because the compiler is able to do vectorization-neutral optimizations within 
code written using intrinsics while it cannot do so when we do manual vectorization through assembly.
On all platforms, vectorization on the BT kernel lowers execution time proportionally but not linearly to the vector 
unit bit-width. This is because there is a small but still computationally significant part of the BT kernel which 
is not vectorizable. By contrast, the MC kernel gains little to nothing from vectorization due to the algorithmic reasons 
discussed earlier. On all platforms autovectorization seems to have a marginal effect 
(Tables~\ref{tab:vpowersave},~\ref{tab:ipowersave}, ~\ref{tab:xpowersave}), indicating inability of the compiler to 
break dependencies that prevent vectorization. The one notable exception is the vectorization of the BT algorithm by 
the ICC compiler on the Xeon Phi, which is nearly as effective as manual vectorization.

For Intel and Viridis, vectorization has little effect on power consumption, increasing it slightly in some cases. 
An interesting observation, which we are still investigating, is that vectorization on Phi actually reduces 
power consumption, see Table~\ref{tab:xpowersave}.

\section{Fair Comparison of Servers and Microservers} \label{sec:experimentsandresults2}   

In this section we are using our workload-specific and platform-independent metrics to directly
compare our three server platforms. We refer to results provided in detail in Table~\ref{tab:calxedaintelmcbtcomparisonsingle}. More detailed results are provided in Tables~\ref{tab:vpowersave}, ~\ref{tab:ipowersave}, and ~\ref{tab:xpowersave} in the appendices. We perform 
comparisons using the vectorized or non-vectorized version of each kernel that achieves the fastest time per 
option priced on each platform.
The fastest version of each kernel is also the least energy consuming version, with one exception,
the MC kernel with 1 million iterations, executed on two Sandy Bridge sockets, where the SSE128
vectorized version consumes less energy than the autovectorized version, albeit being slightly slower. 
We report results from three sets of experiments with each kernel on each platform, where we vary the
number of iterations in each kernel, reflecting different margins of error from the ideal
analytical (Black Scholes) pricing solution for each kernel, which can be traded for execution
speed in realistic market scenarios. We also report PRE-VRM power measurements on all platforms.

\subsection{Monte Carlo Pricing}

A single Viridis microsever underperforms a single Sandy Bridge socket in the MC kernel by one order of
magnitude, with both running in performance mode. While this result can be traced to the many
microarchitectural differences between the two platforms, our experiments suggest that the lack
of efficient hardware implementation of transcendental functions on the ARM Cortex A9 processors is the
critical culprit. That said, the Sandy Bridge socket expends almost six times as much power as the single microserver. 

Taking the MC kernel with one million iterations as an example, we observe that counterbalancing low performance with low lower consumption yields a modest 18\% energy loss to price a single option on Viridis.  
Notably, scaling out the Viridis microservers achieves a near ideal fifteen-fold speedup in time per option priced, with sustained energy consumed per option, as power also scales linearly to the number of microserver nodes.
Scaling out from one to two Sandy Bridge sockets yields a non-ideal speedup of 1.63, while power consumption doubles.  This narrows the gap in energy consumed per option to under 3\% between the Sandy Bridge (better) and the Viridis (worse), the latter with scale out implemented. Sixteen ARM microserver nodes actually outperform
two Sandy Bridge sockets in time per option, but incur an additional power tax of approximately 15 Watts. 

Running the kernel with more or less than a million iterations indicates similar
performance and energy-efficiency trends. The gap in energy per option priced between the Viridis microservers
and the Sandy Bridge in scale out setup is a marginal 1.4\%.

\begin{table}[htbp]
\footnotesize
\centering
\caption{Fastest S/Opt profiles for standalone kernel experiments}
\begin{tabular}{lrrrrr} 
\toprule
\multicolumn{1}{l}{\bf Kernel and} & 
\multicolumn{1}{c}{\bf N}         & 
\multicolumn{1}{c}{\bf VEC TYPE}         & 
\multicolumn{1}{c}{\bf PRE-VRM}   & 
\multicolumn{1}{c}{\bf S/Opt}  & 
\multicolumn{1}{c}{\bf J/Opt}    \\
\multicolumn{1}{l}{\bf Platform}  && 
\multicolumn{4}{c}{\bf $\bar{P}$(W)}     
\\ 
\midrule \textbf{MC Viridis(1x4x1)} & 0.5M & INTRINSICS\_NEON128 & 7.137 & 0.053298 & 0.380382 \\
& 1.0M & NEON128 & 7.258 & 0.105496 & 0.765691 \\
& 2.0M & NEON128 & 7.354 & 0.210560 & 1.548440 \\
\midrule \textbf{MC Viridis(16x4x1)} & 0.5M & NEON128 & 101.930 & 0.003761 & 0.383398 \\
& 1.0M & INTRINSICS\_NEON128 & 102.680 & 0.007205 & 0.739765 \\
& 2.0M & NEON128 & 103.100 & 0.014316 & 1.476006 \\
\midrule \textbf{MC Intel(1x8x1)} & 0.5M & AUTOVECT & 44.899 & 0.007138 & 0.320485 \\
& 1.0M & SSE128 & 44.852 & 0.013896 & 0.623238 \\
& 2.0M & SSE128 & 45.078 & 0.027672 & 1.247364 \\
\midrule \textbf{MC Intel(2x8x1)} & 0.5M & SSE128 & 87.814 & 0.004331 & 0.379259 \\
& 1.0M & NOVECT & 90.999 & 0.008274 & 0.753282 \\
& 2.0M & AUTOVECT & 96.691 & 0.016213 & 1.568116 \\
\midrule \textbf{MC Xeon Phi(1x1x1)} & 0.5M & KNC512 & 22.468 & 0.237212 & 5.329578 \\
& 1.0M & INTRINSICS\_KNC512 & 22.510 & 0.473741 & 10.663713 \\
& 2.0M & NOVECT & 22.852 & 0.947407 & 21.650439 \\
\midrule \textbf{MC Xeon Phi(1x60x1)} & 0.5M & KNC512 & 48.690 & 0.004589 & 0.223363 \\
& 1.0M & AUTOVECT & 52.022 & 0.008925 & 0.464319 \\
& 2.0M & AUTOVECT & 54.065 & 0.017400 & 0.940726 \\
\midrule \textbf{MC Xeon Phi(1x60x2)} & 0.5M & NOVECT & 51.821 & 0.003582 & 0.185607 \\
& 1.0M & NOVECT & 55.246 & 0.006720 & 0.371162 \\
& 2.0M & NOVECT & 57.632 & 0.012860 & 0.741166 \\
\midrule \textbf{MC Xeon Phi(1x60x4)} & 0.5M & NOVECT & 55.087 & 0.002967 & 0.163309 \\
& 1.0M & NOVECT & 59.288 & 0.005348 & 0.317089 \\
& 2.0M & NOVECT & 62.804 & 0.010188 & 0.639805 \\
\midrule
\midrule \textbf{BT Viridis(1x4x1)} & 4000 & NEON128 & 6.246 & 0.007143 & 0.044611 \\
& 5000 & NEON128 & 6.317 & 0.011250 & 0.071070 \\
& 7000 & INTRINSICS\_NEON128 & 7.063 & 0.022279 & 0.157352 \\
\midrule \textbf{BT Viridis(16x4x1)} & 4000 & NEON128 & 93.017 & 0.000649 & 0.060332 \\
& 5000 & NEON128 & 93.783 & 0.000973 & 0.091239 \\
& 7000 & INTRINSICS\_NEON128 & 98.650 & 0.001702 & 0.167785 \\
\midrule \textbf{BT Intel(1x8x1)} & 4000 & INITRINSICS\_AVX256 & 40.494 & 0.001302 & 0.052729 \\
& 5000 & INITRINSICS\_AVX256 & 45.754 & 0.002222 & 0.101594 \\
& 7000 & AVX256 & 48.642 & 0.005192 & 0.252550 \\
\midrule \textbf{BT Intel(2x8x1)} & 4000 & AVX256 & 86.159 & 0.000710 & 0.060584 \\
& 5000 & INITRINSICS\_AVX256 & 80.621 & 0.001464 & 0.118039 \\
& 7000 & INITRINSICS\_AVX256 & 95.177 & 0.003192 & 0.303158 \\
\midrule \textbf{BT Xeon Phi(1x1x1)} & 4000 & AUTOVECT & 23.351 & 0.010955 & 0.255816 \\
& 5000 & INTRINSICS\_KNC512 & 22.758 & 0.017146 & 0.390209 \\
& 7000 & AUTOVECT & 23.012 & 0.033172 & 0.763351 \\
\midrule \textbf{BT Xeon Phi(1x60x1)} & 4000 & INTRINSICS\_KNC512 & 23.833 & 0.000387 & 0.009229 \\
& 5000 & INTRINSICS\_KNC512 & 25.417 & 0.000617 & 0.015824 \\
& 7000 & KNC512 & 29.000 & 0.000745 & 0.021609 \\
\midrule \textbf{BT Xeon Phi(1x60x2)} & 4000 & INTRINSICS\_KNC512 & 26.583 & 0.000401 & 0.010666 \\
& 5000 & KNC512 & 29.861 & 0.000466 & 0.013968 \\
& 7000 & KNC512 & 39.238 & 0.000731 & 0.028691 \\
\midrule \textbf{BT Xeon Phi(1x60x4)} & 4000 & INTRINSICS\_KNC512 & 24.167 & 0.000544 & 0.013149 \\
& 5000 & AUTOVECT & 26.389 & 0.000586 & 0.015522 \\
& 7000 & INTRINSICS\_KNC512 & 30.917 & 0.000891 & 0.027531 \\
\bottomrule
\end{tabular}
\label{tab:calxedaintelmcbtcomparisonsingle} 
\end{table}  

The Xeon Phi server consumes more than two times less energy per option in the Monte Carlo kernel compared to both Viridis and the Sandy Bridge server. Notably, the Xeon Phi does so without employing vectorization and with a significant boost in performance and energy-efficiency from Hyperthreading on each core. Taking
the case with one million iterations as an example (other cases behave similarly), the Phi
with all cores and threads activated provides a seemingly modest 34\% performance improvement over the
Viridis (more compared to Sandy Bridge) but offers a significantly better power-efficiency proposition with 60 cores and 240 threads burning under 60 Watt, when the scaled out Viridis microservers offers 64 cores at 
just over 100 Watt. Note that a single Phi core is two to three times slower than one Sandy Bridge core and 
marginally slower with one ARM Cortex A9 core in terms of performance.

\paragraph{Conclusion:} For a real-time option pricing workload using Monte Carlo methods, the Xeon Phi is the best performing and most energy-efficient option. A scaled out ARM microserver is in par with a heavyweight Sandy Bridge server in terms of energy cost, but can provide higher performance.

\subsection{Binomial Tree}

The lesser dependence of the BT kernel on transcendental functions becomes apparent from single-node
experiments: A single microserver is 5.5 times slower than a single Sandy Bridge socket (vis. 7.5 times in 
the MC kernel). A single Phi core
is 2.5 times faster than a single ARM Cortex A9 core thanks to the higher quality vectorization of the
BT kernel on the Phi and the more powerful SIMD set available on the Phi, compared to the A9.  

Scaling out the microserver yields dramatically lower energy consumption per option priced than the Sandy 
Bridge server. For example, running the BT kernel with 7000 iterations consumes 81\% less Joules per option
on microservers compared to the Sandy Bridge.  Sixteen microservers also outperform two Sandy Bridge sockets
by as much as 46\%. Note that in scale out mode the power consumption of both the Viridis and the Sandy
Bridge servers are comparable. This surprising result -considering the difference in single-core 
performance between the two platforms- can be explained by workload
effects. The Viridis server offers the fact that the 
BT model has two imbalanced computational phases. The setup phase has linear 
complexity and involves transcendental functions,
which are faster on Intel hardware.  The second phase is the tree scanning phase, which has quadratic 
complexity, and uses only floating point addition and multiplication operations. These operations execute 
natively on both platforms. While the Sandy Bridge has an advantage over the ARM on the initialization
phase, this advantage diminishes in the computation phase and is marginalized as the number of iterations
of the BT kernel grows.

The Xeon Phi offers once again an excellent power-efficiency proposition. Compared to the Viridis microservers in scale out setup, the Phi achieves twice the performance, at a third of the power budget, for a 
sixfold reduction of energy per option priced. 

\paragraph{Conclusion:} For a real-time option pricing workload using  Binomial Tree methods, the Xeon Phi is the best performing and most energy-efficient option. A scaled out ARM microserver is significantly faster
and more energy-efficient than an equivalent, from a power perspective, Sandy Bridge server.
 
\subsection{QoS Discussion}   

In a live market situation the question arises whether option pricing 
can keep pace with the rate of arrival of new stock prices.
We use the QoS metric defined in Section~\ref{sec:Experimental Setup and Measurement Methodology} to consider
the implications of dynamic deadlines set by market conditions on the option pricing workload.
We define an \textit{iso-QoS} comparison method, where we compare the performance, power and energy of
different platforms, with setups that achieve the same QoS for a given workload. 
One can conduct similar exercises using iso-metrics that equate power, performance, energy consumption,
or combined efficiency metrics.

\begin{figure}[htbp]
  \centering
  \includegraphics[width=0.6\textwidth]{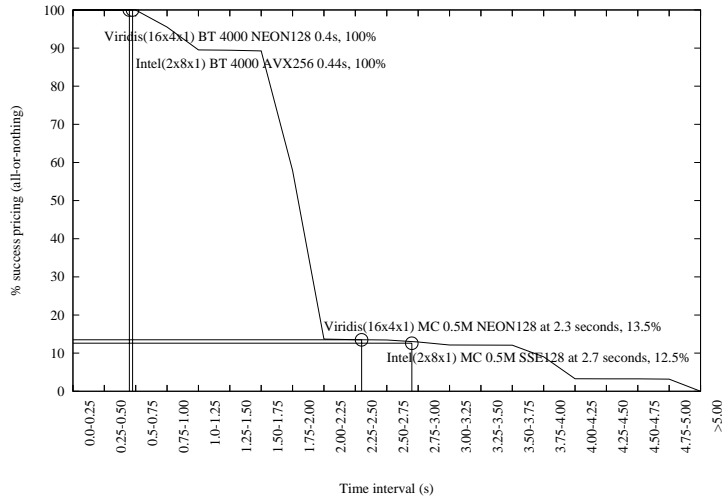}
  \caption{All-or-nothing pricing vs. stock price update intervals}
  \label{fig:cumulativepricefreq}
\end{figure}

A new stock price arrival triggers a computation of all 
available option contracts. The most stringent QoS requirement is 
``all-or-nothing": the pricing of all options must complete 
before another stock price update, otherwise any computation performed before the
deadline is discarded, having wasted time and energy. 

Indicatively, Figure~\ref{fig:cumulativepricefreq} shows the 
percentage of successful all-or-nothing option pricing computations as a function of 
the cumulative number of price updates, sorted into bins of 0.25 
second intervals. The data for our experiments are taken 
from a trading session of 6.5 hours where 10,156 price updates occurred 
for the Facebook stock, producing the cumulative profile shown in the figure. 

Nonetheless, there are trading scenarios where the financial user does 
not require this all-or-nothing approach. For example a user may 
prefer to price a subset of options, such as short-term or long-term 
expiring ones, for which there is financial interest. Moreover, the 
all-or-nothing approach disregards the fact that some options were 
indeed correctly priced and they can be used for trading. For these 
reasons we relax this requirement and defined our 
QoS metric as the percentage of successfully priced option contracts 
over the total number of option contract pricings.

Figure~\ref{fig:cumulativepricefreq} shows the QoS profile of the Viridis and Sandy Bridge platforms
with two scenarios of Monte Carlo (half a million iterations) and Binomial Tree (4000 iterations).
Both platforms run in performance mode and are scaled to the maximum number of cores available.
The figure shows the performance gap between options pricing on the two platforms.
The microserver is able to price all options using Monte Carlo slightly faster ($[2.25-2.50]$ seconds time bin), while the Sandy Bridge is slower ($[2.50-2.75]$ seconds bin) but both platforms exhibit low QoS
(13.5\% and 12.5\% respectively), which may be unacceptable for an end user. 
On the other hand, both platforms exhibit 100\% QoS with the Binomial Tree kernel, where the Viridis 
microservers also reduce total energy consumption by as much as 55\%.

\section{Related Work}
\label{sec:related}

Recent work related to ours explores the performance and power consumption of low-power ARM processor
models that attempt to enter the server~\cite{Tudor:2013:UEC:2465529.2465553,Ou:2012:ECA:2310096.2310142}
and HPC markets~\cite{Rajovic:2013:SCC:2503210.2503281}. Our study targets a different domain, 
real-time financial analytics, and provides a comparison between fully integrated ARM-based microservers in scale-out configurations,
against high-end Intel servers. We also provide further insight on the viability of the
ARM ecosystem for datacenters and high-performance computing, using new platform-independent metrics
and a comprehensive method to ensure fair comparison. The work of Blem et 
al~\cite{Blem:2013:PSR:2495252.2495491} is perhaps the closest to ours. It explores the performance and power 
consumption of several ARM and Intel processors. However, their study focuses on the energy and performance 
implications of ISA choices on the different processors, rather than the use of the processors at scale
or their deployment in specific application domains.
Anwar et al~\cite{Anwar:2014} study microserver performance with Hadoop workloads.
Our work advances the state of knowledge in microservers, by exploring the performance and energy 
implications of scaling their resources up and out and by placing microservers on a 
fair comparison basis against both general-purpose servers and accelerators targeting the 
HPC market.

Also related to our contribution is work that measures the energy-efficiency of specific algorithms
and numerically intensive kernels. 
Alonso et al~\cite{DBLP:journals/ife/AlonsoDMQ14} model the
power and energy of a specific task-parallel implementation of Cholesky factorization.
Dongarra et al~\cite{6382829} explore the energy footprint of dense numerical linear
algebra libraries on multicore systems.
Korthikanti et al~~\cite{conf/spaa/KorthikantiA10} analyzed the energy scalability of parallel
algorithms on shared-memory multicore architectures, focusing on the sensitivity of the 
the algorithms to the critical path length, the power consumption of instruction execution and 
the power consumption of memory accesses.
Within the European PRACE project, a performance 
specific to 2D stencil applications on multi-core parallel machines was
derived~\cite{HouPrace:2012}. Kozin~\cite{kozin:2011} analyzed the execution of several computational chemistry codes on Nehalem 
servers and on a heterogeneous system composed of one Nehalem server and one 
Tesla GPU system. They reported that idle power was a significant 
component of overall power consumption. Our work differs in that it explores a dynamic workload
with real-time execution requirements and a mixture of event-driven and data parallelism.

\typeout{>>>> Processing the Conclusions section}

\section{Conclusions} \label{sec:conclusions}

In this paper we have presented a fair methodology for comparing 
server platforms for real-time financial analytics. Our methodology
considers performance, energy-efficiency, and QoS impact on energy consumption, hence cost of provisioning these services. 
We have based our methodology on workload-specific but platform-independent metrics, which facilitated
direct performance and cost comparisons. These comparisons will benefit directly datacenter operators 
during hardware procurement and provisioning exercises, but also developers of financial services during
exercises to reduce service cost while sustaining competitive performance and QoS in their products.
Notably, our study used real stock market streaming data and captured the dynamic, event-driven
nature of real-time financial analytics workloads. 

Our results show that microservers based on ARM processors are viable 
contenders to state-of-the-art general-purpose servers. Scaled out 
microserver configurations achieve similar or better performance and similar
or improved energy-efficiency under a fixed power budget. However, microservers
can not match densely packaged many-core accelerators, such as the Xeon Phi.
The introduction of such computational accelerators within the power and
area constraints of microservers is thus a promising area of further investigation.
Extending our methodology and metrics to heterogeneous platforms is also 
appropriate given the increasing presence of such platforms in datacenters.
providing the same quality of service.
One further open area for future research is the exact error behavior of the MC and BT models in
response to reducing their energy consumption by reducing their number of iterations.

\begin{center}
\textbf{Acknowledgments}
\end{center}

The work was supported by the European Commission under its Seventh Framework Programme, grant
number 610509 (NanoStreams). This work was also supported by the UK Engineering and Physical Sciences 
Research Council, under grant agreements EP/K017594/1, EP/L000055/1 and EP/L004232/1.


\typeout{>>>> Creating the references now}

\bibliographystyle{wileyj}
\bibliography{nanostreams_submission}

\typeout{****  Finished creating the references now - End of document}

\newpage
\appendix
\section{Appendix} \label{sec:appendix}
%
%
\appendix
\begin{table}[ht]
\centering
\caption{Viridis S/Opt and J/Opt metrics, using PRE-VRM power}
\begin{tabular}{lrrrrrr}
\toprule
& \multicolumn{3}{c}{\bf Performance}   & \multicolumn{3}{c}{\bf Powersave} \\ 
\cmidrule{2-7}
& \multicolumn{1}{c}{\bf $\bar{P} (W)$} & \multicolumn{1}{c}{\bf S/Opt} & \multicolumn{1}{c}{\bf J/Opt}  
& \multicolumn{1}{c}{\bf $\bar{P} (W)$} & \multicolumn{1}{c}{\bf S/Opt} & \multicolumn{1}{c}{\bf J/Opt} \\ 
\midrule

\textbf{Viridis(1x4x1)} &&&&&\\
\textbf{MC 1M}   
NOVECT & 7.19922&0.11472&0.82590&4.33896&0.90668&3.93405\\
AUTOVECT & 7.27235&0.11522&0.83793&4.34735&0.94811&4.12176\\
NEON128 & 7.25806&0.10550&0.76570&4.31920&0.83421&3.60314\\
INTRINSICS\_NEON128 & 7.23137&0.10810&0.78171&4.32488&0.84103&3.63736\\ \midrule
\textbf{BT 5000} 
NOVECT & 7.44259&0.01584&0.11786&4.28633&0.11307&0.48467\\
AUTOVECT & 7.39630&0.01580&0.11687&4.26875&0.11324&0.48340\\
NEON128 & 6.31710&0.01125&0.07107&4.20724&0.08048&0.33858\\
INTRINSICS\_NEON128 & 7.11229&0.01270&0.09032&4.46305&0.10792&0.48163\\ \midrule
\textbf{Viridis(16x4x1)} &&&&&\\
\textbf{MC 1M}   
NOVECT & 101.00667&0.00820&0.82863&63.17333&0.06537&4.12953\\
AUTOVECT & 102.50000&0.00817&0.83756&62.41000&0.07180&4.48118\\
NEON128 & 102.93333&0.00731&0.75209&62.57000&0.06431&4.02363\\
INTRINSICS\_NEON128 &  102.68000&0.00721&0.73976&85.13000&0.02644&2.25080\\ \midrule
\textbf{BT 5000} 
NOVECT & 103.14333&0.00130&0.13379&63.22667&0.00773&0.48885\\
AUTOVECT & 103.83500&0.00130&0.13469&61.58333&0.00790&0.48618\\
NEON128 & 93.78333&0.00097&0.09124&64.25000&0.00400&0.25707\\
INTRINSICS\_NEON128 &  93.72333&0.00097&0.09118&68.75667&0.00249&0.17106\\ \midrule
\bottomrule
\end{tabular}
\label{tab:vpowersave}
\end{table}
  
\begin{table}[ht]
\centering
\caption{Intel Sandy Bridge S/Opt and J/Opt metrics, using PRE-VRM power}
\begin{tabular}{lrrrrrr}
\toprule
& \multicolumn{3}{c}{\bf Performance}   & \multicolumn{3}{c}{\bf Powersave} \\ 
\cmidrule{2-7}
& \multicolumn{1}{c}{\bf $\bar{P} (W)$} & \multicolumn{1}{c}{\bf S/Opt} & \multicolumn{1}{c}{\bf J/Opt}  
& \multicolumn{1}{c}{\bf $\bar{P} (W)$} & \multicolumn{1}{c}{\bf S/Opt} & \multicolumn{1}{c}{\bf J/Opt} \\ 
\midrule
\textbf{Intel(1x8x1)} &&&&&\\
\textbf{MC 1M}
NOVECT & 45.04794&0.014325&0.645297&28.689189&0.024103&0.691485 \\
AUTOVECT & 45.269019&0.014271&0.646025&28.692122&0.024109&0.69175 \\
SSE128 & 44.85186&0.013896&0.623248&28.579088&0.02335&0.667322 \\
AVX256 & 45.26695&0.024919&1.128025&28.721541&0.042054&1.207863 \\
INTRINSICS\_AVX256 &  45.441663&0.0253&1.149656&28.739781&0.043073&1.237895 \\
\textbf{BT 5000} 
NOVECT & 43.27172&0.00584&0.252717&27.78212&0.009733&0.270402\\
AUTOVECT & 43.293392&0.005838&0.252739&27.798964&0.009732&0.27053\\
SSE128 & 44.113263&0.003894&0.171773&28.989414&0.006164&0.178696\\
AVX256 & 45.450668&0.002275&0.1034&30.116217&0.00357&0.107502\\
INTRINSICS\_AVX256 &  45.754212&0.002222&0.101684&29.627705&0.003568&0.10572\\ \midrule
\textbf{Intel(2x8x1)} &&&&&\\
\textbf{MC 1M}
NOVECT & 90.999305&0.008274&0.7529&57.925196&0.01482&0.858431 \\
AUTOVECT & 90.048869&0.008453&0.761147&54.843581&0.014646&0.803252 \\
SSE128 & 84.694158&0.008493&0.719341&60.652193&0.012971&0.786716 \\
AVX256 & 88.619245&0.014214&1.259592&57.493955&0.024968&1.435523 \\
INTRINSICS\_AVX256 &  89.048027&0.015029&1.338341&57.128128&0.026699&1.525274 \\
\textbf{BT 5000} 
NOVECT & 85.380722&0.00364&0.310779&55.24572&0.006228&0.344074\\
AUTOVECT & 81.84787&0.00368&0.301215&69.158565&0.005629&0.389319\\
SSE128 & 79.325308&0.002655&0.210598&55.302365&0.003908&0.216115\\
AVX256 & 79.732068&0.001525&0.121553&57.545024&0.002342&0.134773\\
INTRINSICS\_AVX256 &  80.620701&0.001464&0.118034&59.139075&0.002166&0.128069\\ \midrule

\bottomrule
\end{tabular}
\label{tab:ipowersave}
\end{table}

\begin{table}[ht]
\centering
\caption{Xeon Phi Joules, S/Opt and J/Opt metrics using PRE-VRM power}
\begin{tabular}{lrrrrrr}
\toprule
& \multicolumn{3}{c}{\bf Performance}   & \multicolumn{3}{c}{\bf Powersave} \\ 
\cmidrule{2-7}
& \multicolumn{1}{c}{\bf $\bar{P} (W)$} & \multicolumn{1}{c}{\bf S/Opt} & \multicolumn{1}{c}{\bf J/Opt}  
& \multicolumn{1}{c}{\bf $\bar{P} (W)$} & \multicolumn{1}{c}{\bf S/Opt} & \multicolumn{1}{c}{\bf J/Opt} \\ 
\midrule
\textbf{Xeon Phi(1x1x1)} &&&&&\\
\textbf{MC 1M} 
NOVECT & 22.97195&0.4742&10.89323&17.80396&0.59137&10.52876\\
AUTOVECT & 22.4946&0.47401&10.66273&16.3638&0.59208&9.68871\\
KNC512 & 22.53165&0.47481&10.69822&16.64192&0.59276&9.86468 \\
INTRINSICS\_KNC512 & 22.50958&0.47374&10.66371&16.36761&0.59237&9.69560 \\ \midrule
\textbf{BT 5000} 
NOVECT& 22.99814&0.10305&2.36986&17.43763&0.12869&2.24396\\
AUTOVECT & 23.09841&0.01720&0.39733&17.78937&0.02139&0.38046\\
KNC512& 22.71311&0.01716&0.38979&17.87900&0.02139&0.38242\\
INTRINSICS\_KNC512 & 22.75758&0.01715&0.39021&17.87283&0.02139&0.38232\\ \midrule
\textbf{Xeon Phi(1x60x1)} &&&&&\\
\textbf{MC 1M} 
NOVECT& 49.44860&0.00919&0.45453&41.95386&0.01149&0.48213\\
AUTOVECT & 52.02175&0.00893&0.46432&42.38235&0.01154&0.48888\\
KNC512& 51.94204&0.00893&0.46403&42.22966&0.01154&0.48716\\
INTRINSICS\_KNC512 & 51.52656&0.00913&0.47059&42.82180&0.01118&0.47882\\
\textbf{BT 5000} 
NOVECT& 47.44445&0.00208&0.09861&40.55693&0.00254&0.10293\\
AUTOVECT & 26.08333&0.00073&0.01896&20.16667&0.00078&0.01570\\
KNC512& 25.91667&0.00062&0.01603&19.94445&0.00074&0.01479\\
INTRINSICS\_KNC512 & 25.41667&0.00062&0.01568&20.58333&0.00077&0.01574\\ \midrule
\textbf{Xeon Phi(1x60x2)} &&&&&\\
\textbf{MC 1M} 
NOVECT& 55.24644&0.00672&0.37128&47.19607&0.00811&0.38260\\
AUTOVECT & 57.22312&0.00678&0.38811&46.85469&0.00827&0.38760\\
KNC512& 58.64738&0.00675&0.39592&47.04584&0.00814&0.38304\\
INTRINSICS\_KNC512& 57.09140&0.00715&0.40821&46.20823&0.00849&0.39231\\
\textbf{BT 5000} 
NOVECT& 50.37038&0.00177&0.08908&44.01588&0.00213&0.09375\\
AUTOVECT & 30.80159&0.00062&0.01923&26.54167&0.00076&0.02013\\
KNC512& 29.86111&0.00047&0.01392&27.87500&0.00078&0.02176\\
INTRINSICS\_KNC512& 28.47619&0.00056&0.01590&27.57937&0.00073&0.02014\\ \midrule
\textbf{Xeon Phi(1x60x4)} &&&&&\\
\textbf{MC 1M} 
NOVECT& 59.28792&0.00535&0.31709&49.90982&0.00674&0.33656\\
AUTOVECT & 59.76438&0.00573&0.34251&50.66874&0.00701&0.35528\\
KNC512& 61.00413&0.00575&0.35089&51.90346&0.00672&0.34882\\
INTRINSICS\_KNC512& 62.66650&0.00554&0.34707&51.63727&0.00673&0.34761\\
\textbf{BT 5000} 
NOVECT& 47.09524&0.00148&0.06985&41.64147&0.00160&0.06645\\
AUTOVECT & 26.38889&0.00059&0.01545&20.52778&0.00076&0.01554\\
KNC512& 26.11667&0.00083&0.02171&20.66667&0.00085&0.01755\\
INTRINSICS\_KNC512& 25.33333&0.00060&0.01521&21.05556&0.00090&0.01885\\ \midrule
\bottomrule
\end{tabular}
\label{tab:xpowersave}
\end{table}

\end{document}